\documentclass[useAMS,usenatbib]{mn2e}

 \usepackage{times}
 \usepackage{graphicx}
\def\pcm3{{\rm\thinspace cm^{-3}}}

\def\contcaption{\@conttrue\SFB@caption\@captype}

\def\n_h{{\rm n_{H}}}

\def\NH1{{$N_{\rm HI}~$}}

\def\ga{{\rm\thinspace gauss}}

\def\approxlt{\mathrel{\hbox{\rlap{\lower .5ex \hbox {$\sim$}}
        \raise .15 ex \hbox{$<$}}}}
\def\approxgt{\mathrel{\hbox{\rlap{\lower .5ex \hbox {$\sim$}}
        \raise .15 ex \hbox{$>$}}}}

\def\la{\mathrel{\hbox{\rlap{\hbox{\lower4pt\hbox{$\sim$}}}\hbox{$<$}}
}}
\def\ga{\mathrel{\hbox{\rlap{\hbox{\lower4pt\hbox{$\sim$}}}\hbox{$>$}}
}}
\newbox\grsign \setbox\grsign=\hbox{$>$} \newdimen\grdimen
\grdimen=\ht\grsign
\newbox\simlessbox \newbox\simgreatbox \newbox\simpropbox
\setbox\simgreatbox=\hbox{\raise.5ex\hbox{$>$}\llap
     {\lower.5ex\hbox{$\sim$}}}\ht1=\grdimen\dp1=0pt
\setbox\simlessbox=\hbox{\raise.5ex\hbox{$<$}\llap
     {\lower.5ex\hbox{$\sim$}}}\ht2=\grdimen\dp2=0pt
\setbox\simpropbox=\hbox{\raise.5ex\hbox{$\propto$}\llap
     {\lower.5ex\hbox{$\sim$}}}\ht2=\grdimen\dp2=0pt
\def\simgreat{\mathrel{\copy\simgreatbox}}
\def\simless{\mathrel{\copy\simlessbox}}

\title[Open cluster white dwarfs]{Further investigation of white dwarfs in the open clusters NGC\,2287 and NGC\,3532\thanks{}}
\author[Dobbie, Day-Jones, Williams, Casewell, Burleigh, Lodieu, Parker \& Baxter]{P. D. Dobbie$^{1,2}$\thanks{E-mail:pdd@aao.gov.au}, A. Day-Jones$^{3}$,  K.A. Williams$^{4}$, S.L.Casewell$^{5}$, M.R.Burleigh$^{5}$, 
\newauthor N. Lodieu$^{6,7}$, Q.A. Parker$^{2,8}$, R. Baxter$^{8}$ \\
$^{1}$School of Mathematics \& Physics, University of Tasmania, Hobart, TAS, 7001, Australia \\
$^{2}$Australian Astronomical Observatory, PO Box 296, Epping, NSW, 1710, Australia \\
$^{3}$Dept. de Astronomia, Universidad de Chile, Camino del Observatorio 1515, Santiago, Chile \\
$^{4}$Dept. of Physics \& Astronomy, Texas A \& M University-Commerce, PO Box 3011, Commerce, TX 75429, USA\\
$^{5}$Dept. of Physics \& Astronomy, University of Leicester, Leicester, UK, LE1 7RH \\
$^{6}$Instituto de Astrof\'isica de Canarias, V\'ia L\'actea s/n, E-38200 La Laguna, Tenerife, Spain \\
$^{7}$Departmento de Astrof\'isica, Universidad de La Laguna, E-38205 La Laguna, Tenerife, Spain \\
$^{8}$Dept. of Physics \& Astronomy, Macquarie University, NSW, 2109, Australia
}
\begin{document}

\date{Accepted Aye; Received \today{}; in original form of \today{}}

\pagerange{\pageref{firstpage}--\pageref{lastpage}} \pubyear{2010}

\maketitle

\label{firstpage}

\begin{abstract}

We report the results of a CCD imaging survey, complimented by astrometric and spectroscopic follow-up studies, that aims to probe the fate of heavy-weight intermediate mass stars by unearthing new, faint, white dwarf members of the rich, nearby, intermediate age open clusters NGC\,3532 and NGC\,2287. We identify a total of four white dwarfs with distances, proper motions and cooling times which can be reconciled with membership of these populations. We find that WD\,J0643-203 in NGC\,2287, with an estimated mass of $M$=1.02-1.16M$_{\odot}$, is potentially the most massive white dwarf so far identified within an open cluster. Guided by the predictions of modern theoretical models of the late-stage evolution of heavy-weight intermediate mass stars, we conclude that there is a distinct possibility it has a core composed of O and Ne. We also determine that despite the cooling times of the three new white dwarfs in NGC\,3532 and the previously known degenerate member NGC\,3532-10 spanning $\sim$90Myr, they all have remarkably similar masses ($M$$\sim$0.9-1M$_{\odot}$). This is fully consistent with the results from our prior work on a heterogeneous sample of $\sim$50 white dwarfs from 12 stellar populations, on the basis of which we argued that the stellar initial mass-final mass relation is less steep at $M$$_{\rm init}$$>$4M$_{\odot}$ than in the adjacent lower initial mass regime. This change in the gradient of the relation could account for the secondary peak observed in the mass distribution of the field white dwarf population and mitigate the need to invoke close binary evolution to explain its existence. Spectroscopic investigation of numerous additional candidate white dwarf members of NGC\,3532 unearthed by a recent independent study would be useful to confirm (or otherwise) these conclusions.

\end{abstract}


%
%
\section{Introduction}
\label{ngc3532:intro}

Only H, He and Li are believed to have been created in significant quantities during the earliest stages in the evolution of the Universe 
\citep[e.g.][]{wagoner67}. Heavier elements, including the two main constituents of the Earth's atmosphere, O and N, are manufactured by 
stars \citep[][] {burbidge57}. These species are injected into the interstellar medium (ISM) of our Galaxy and others primarily during the
latter phases in the stellar lifecycle when stars, depending on their mass, evolve either towards a supernova explosion or up the asymptotic 
giant branch (AGB) and through the planetary nebula phase. Subsequently, they are mixed into the ambient gas \citep{scalo04} from which they 
will be incorporated into future generations of stars and planets.

A detailed understanding of how stellar populations build up and modify the metallicity of the ISM over time is essential to fully comprehend 
the formation and the evolution of galaxies. An important aspect of this is knowledge of the amount and the composition of the gas a star of 
a particular initial mass (and metallicity) will return to the ISM during it's lifetime. This is intimately linked to it's final evolutionary 
state. Most low and intermediate mass stars ($M$$_{\rm init}$$\simless$5-6M$_{\odot}$) are anticipated to end their lives as either He or CO white 
dwarfs while massive objects ($M$$_{\rm init}$$\simgreat$10M$_{\odot}$) are expected to perish in Fe core-collapse Type II supernovae explosions 
(SNe II). The fate of the heaviest intermediate mass stars (IMS) which lie in the intervening mass range, however, remains very uncertain 
\citep[e.g.][]{siess06}, despite these being comparable in number to the massive objects.
Stellar evolutionary models suggest that the C in a partially degenerate core of an aged heavy-weight IMS ignites when it achieves $M$$\simgreat
$1.05M$_{\odot}$ and burns to O and Ne \citep{nomoto84,nomoto87}, before the star enters a phase of super-AGB evolution. The very high temperature 
environment at the bottom of the convective envelopes of super-AGB stars favours the production of isotopes such as $^{7}$Li, $^{14}$N, $^{13}$C and
$^{26}$Al \citep[][]{siess10}. If this chemically enriched envelope gas is removed before the core attains $M$$\sim$1.37M$_{\odot}$, the star likely 
ends life as an ultra-massive ($M$$\sim$1.1-1.35M$_{\odot}$) ONe white dwarf \citep[e.g][]{ritossa96,garcia97}. However, if the degenerate interior 
reaches $M$$\sim$1.37M$_{\odot}$ then electron captures onto $^{24}$Mg and $^{20}$Ne will lead to a core collapse and a weak type II electron capture 
supernova explosion \citep[ECSNe;][]{nomoto84}. The temperatures and particle densities inherent to ECSNe are anticipated to generate a somewhat 
more exotic cocktail of elements that also includes $^{64}$Zn, $^{70}$Ge, $^{90}$Zr and possibly heavy r-process elements \citep[A$>$130;][]{qian08},
but which is relatively poor in $\alpha$-process elements and Fe \citep{wanajo09}.

\begin{table*}
\begin{minipage}{161mm}
\begin{center}
\caption{Details of the four white dwarf members of NGC\,2516 derived from FORS1 (top) and FORS2 (bottom) observations.  The tabulated 
spectroscopic signal-to-noise estimates correspond to per resolution element over the range $\lambda$=4150-4300\AA. Masses and cooling times for 
each star have been estimated using the mixed CO core composition ``thick H-layer'' evolutionary calculations of 
the Montreal Group (e.g. Fontaine, Brassard \& Bergeron 2001). The quoted limits on the near-IR magnitudes of the white dwarfs correspond to 3$\sigma$ detections in the images.}

\label{wdmass2}
\begin{tabular}{cccccccccc}
\hline
 ID & S/N &$T_{\rm eff}$(K)$^{*}$ & log $g$$^{*}$ &  $V$$^{\dagger}$  & $J$ & $K_{\rm S}$ & M$_{V}$ & M(M$_{\odot}$)  &  $\tau_{c}$ (Myr) \\
\hline
  NGC\,2516-1 & 65 & $29354^{+329}_{-341}$ & $8.48^{+0.05}_{-0.04}$ & 19.18$\pm$0.10 & $>$19.6 & $>$18.0 & $10.80^{+0.13}_{-0.13}$ & $0.93\pm0.04$  & $49^{+11}_{-11}$ \\ \\
  NGC\,2516-2 & 75 & $34913^{+533}_{-400}$ & $8.53^{+0.06}_{-0.06}$ & 19.27$\pm$0.07 & 19.08$\pm$0.19 & $>$18.0 & $10.56^{+0.13}_{-0.13}$ & $0.97\pm0.04$  & $23^{+8}_{-7}$\\ \\
  NGC\,2516-3 & 130 & $28708^{+166}_{-176}$ & $8.49^{+0.03}_{-0.03}$ & 19.46$\pm$0.13 & $>$19.6 & $>$18.0 & $10.87^{+0.13}_{-0.13}$ & $0.94\pm0.04$  & $55^{+12}_{-11}$  \\ \\
  NGC\,2516-5 & 150 & $31844^{+173}_{-174}$ & $8.54^{+0.03}_{-0.03}$ & 19.24$\pm$0.14 & $>$19.6 & $>$18.0 & $10.74^{+0.14}_{-0.14}$ & $0.97\pm0.04$  & $40^{+10}_{-9}$   \\
\hline
 NGC\,2516-1 & 75 & $29191^{+312}_{-321}$ & $8.50^{+0.04}_{-0.04}$ & - & - & - & $10.85^{+0.13}_{-0.13}$ & $0.94\pm0.04$  & $53^{+12}_{-11}$  \\
\hline
\end{tabular}
\end{center}
$^{*}$ Formal fitting uncertainties \\
$^{\dagger}$ As reported in \citet{koester96}
\end{minipage}
\end{table*}

Unfortunately, the treatment of key physical processes within these calculations remains rather rudimentary. This limits the capacity of stellar 
evolutionary theory to make firm predictions regarding the fate of stars in this mass range. For example, the details of how convective 
mixing is implemented in the models, especially during the He core burning phase, can systematically shift the initial mass range of super-AGB 
stars by $\sim$2M$_{\odot}$ \citep[e.g.][]{gilpons07}. It remains an open question as to whether during the super-AGB phase the atmospheres 
are cool enough to form dust \citep[][]{poelarends08}. Dust appears to be essential for the efficient transfer of net outward momentum from the 
radiation field to the envelope gas, at least in thermally-pulsing AGB stars \citep[e.g.][]{wachter02}, so it is still uncertain if
single stars in this mass range can evolve into ultra-massive white dwarfs (UMWDs) at all.

Investigations of field UMWDs have been unable to provide unambiguous evidence in support of an evolutionary link with heavy-weight IMSs, perhaps 
because this sizeable population is heterogeneous in origin. For example, \cite{liebert05a} found that the spatial distribution of the highest 
mass white dwarfs detected in the extreme ultraviolet (EUV) sky surveys, including 10 with $M$$\ga$1.1M$_{\odot}$, is remarkably similar to that 
of the young stellar populations in the vicinity of Gould's Belt. It has been argued that, given their space motions and cooling times, the UMWDs 
GD\,50 and PG\,0136+251 could be associated with the young Pleiades open cluster \citep{dobbie06b}. In contrast, however, the empirical white dwarf
mass distribution, when compared to the output of traditional population synthesis calculations, suggests that the majority of UMWDs may be produced
through the merging of the lower mass components of primordial close binary systems, at least at the higher galactic latitudes\citep[][]{yuan92, 
liebert05a}. 

The study of white dwarfs that are demonstrably bound members of young open clusters can provide better insight into the fate of heavy-weight IMSs.
This is because these populations are co-eval and have comparatively well determined ages, making it possible to place tighter constraints on the prior 
evolution of their degenerate members. With ready access to large telescopes, the number of high mass white dwarfs known to be members of open clusters 
has increased substantially during the last few years \citep[$\sim$30 degenerates with $M$$>$0.8M$_{\odot}$ in $\sim$10 populations, e.g.][]{williams04b,
williams09, kalirai05, dobbie04}. However, there remain very few, if any, known cluster white dwarfs with $M$$\simgreat$1.05M$_{\odot}$. Amongst the most 
promising 
candidates are LB\,1497 \citep[Pleiades; ][]{bergeron92}, KRR\,24 \citep[NGC\,2099; ][]{kalirai05} and four objects in NGC\,2168 \citep{williams09}, but
their estimated masses are either borderline or have large uncertainties due to the rather modest quality of the spectroscopy used to determine their 
fundamental parameters. It is unclear if this lack of compelling candidate cluster UMWDs is a manifestation of a real deficit or merely down to counting 
statistics. At present, NGC\,2168 is by far the richest young population in which the white dwarf cooling sequence has been spectroscopically studied 
down to its termination. Extrapolation of the cluster mass function beyond the main sequence turn-off mass suggests that the central regions of NGC\,2168 
should contain $\sim$18 white dwarf members \citep{kalirai03}. \cite{williams09} identified 12-14 possible degenerate members from their $UBV$ imaging 
of the cluster core which is marginally consistent with this prediction (P$\sim$0.09 for 12 objects), especially considering that a non-zero proportion 
of the cluster white dwarfs will be hidden from view by binary companions in earlier evolutionary phases \citep[e.g.][]{williams04a}. 

In light of this we have undertaken a new survey of two open clusters, NGC\,3532 and NGC\,2287, that aims to identify and characterise the white dwarf 
remnants of their recently deceased heavy-weight IMS members. Both targets are rich, nearby, young/intermediate age populations \citep[$\tau
$=300$\pm$25Myr and $\tau$=243$\pm$40Myr, respectively,][]{meynet93,koester93,harris93, kharchenko05a,sharma06} which, as outlined in \cite[][]{dobbie09a}, 
have been previously searched photographically for white dwarfs \citep[e.g.][]{romanishin80, reimers89}. However, none of the six photometric candidates 
subsequently confirmed through spectroscopy as white dwarf members of these two populations \citep[][]{koester81, koester93, dobbie09a} appear to have formed 
from progenitors with $M$$\simgreat$5M$_{\odot}$, despite both clusters being relatively rich. It is possible the remnants of these stars have remained undetected
until now because they lie beyond the detection limits of the photographic plates \cite[e.g.][]{clem11}. Indeed, the total mass of NGC\,3532 is estimated to be 
$M$$_{\rm tot}$$>$2000M$_{\odot}$ \citep{fernandez80}. This is approximately twice that of the 150Myr old cluster, NGC\,2516 \citep{jeffries01b}, the central regions of which harbour four massive white dwarfs descended from stars with $M$$\simgreat$5M$_{\odot}$ \citep{koester96}. On the assumption that the initial mass functions of these populations were of a similar power law form \citep[e.g.][]{salpeter55}, it could be expected that 4$\pm$2 degenerate remnants from stars with $M$$\simgreat$5.5M$_{\odot}$ reside within NGC\,3532 and approximately half this number within NGC\,2287.

In subsequent sections we outline our new, more sensitive charged coupled device (CCD) based survey of these clusters and detail our initial selection of candidate white dwarf members. We describe our spectroscopic follow-up observations of these sources, the analysis of the spectral datasets and the determination of the fundamental parameters of the new white dwarfs uncovered. In Section\ \ref{ngc3532:WDFP} we estimate the distances and proper motions of the degenerates and compare these to the properties of their proposed host clusters. For those white dwarfs we find to be probable members of NGC\,3532 and NGC\,2287, we derive progenitor masses and conclude by discussing them in the context of stellar evolution and the stellar initial mass-final mass relation. First, however, we briefly re-visit the four high mass white dwarf members of NGC\,2516, which have traditionally been used to define the upper end of the IFMR. We use our new and improved spectroscopy to place tighter limits on their fundamential parameters and inferred progenitor masses.

\begin{figure}
\includegraphics[angle=0,width=\linewidth]{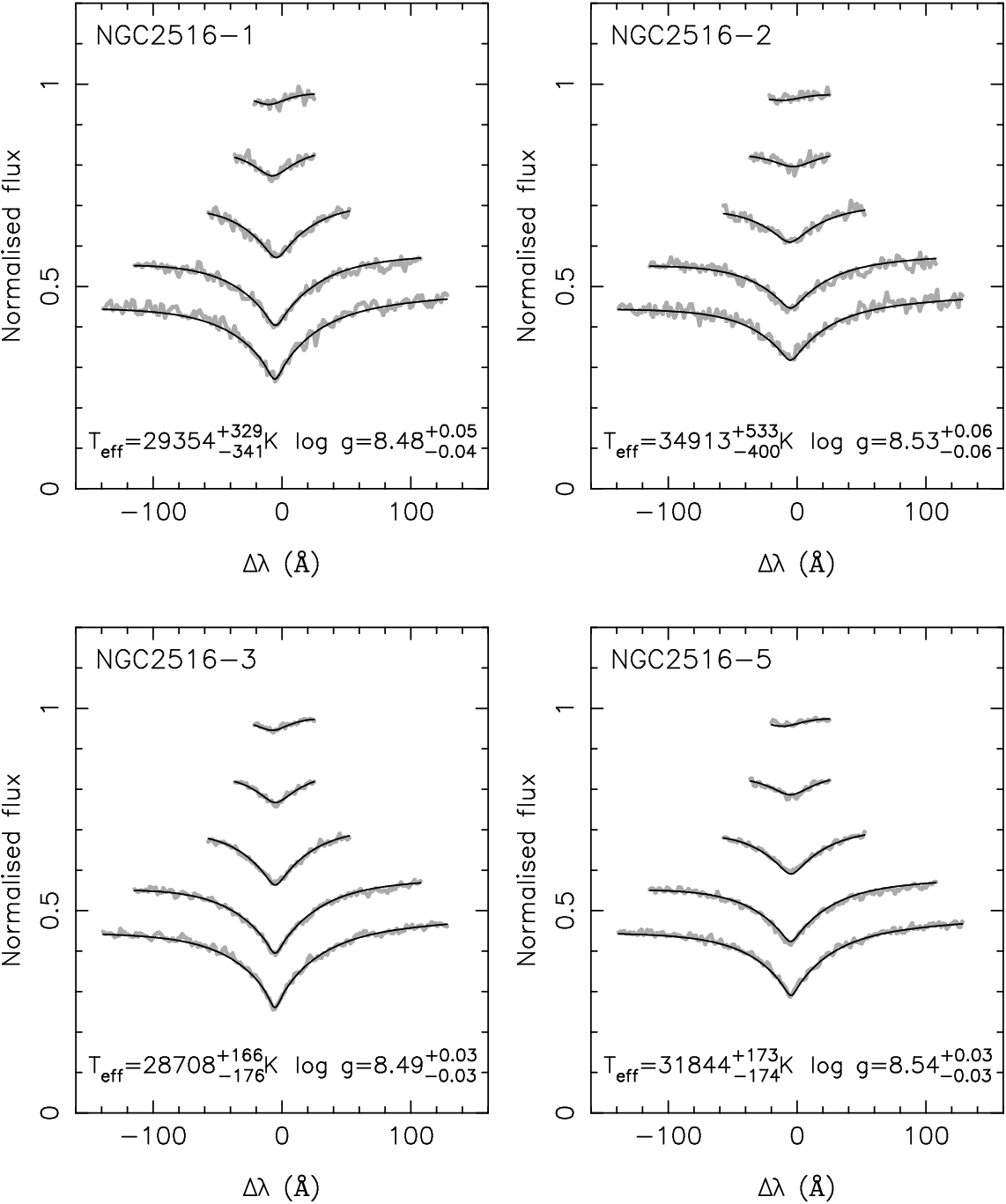}
\caption{The results of our fitting of synthetic profiles (thin black lines) to the observed Balmer lines, H-$\beta$ to H-8
, of the four white dwarf candidate 
members of NGC\,2516 (thick grey lines). The flux$_{\lambda}$ units are arbitrary.}
   \label{LINES}
\end{figure}

\section{The NGC\,2516 white dwarfs revisited}

The open cluster NGC\,2516 was recognised several decades ago as an attractive target for investigating the form of the upper end of the IFMR \citep[e.g][]{reimers82}. It is young (e.g. 140Myr, \citealp{meynet93}; 158Myrs, \citealp{sung02}; 120Myrs, \citealp{kharchenko05a}), comparatively nearby, (e.g. 430$^{+80}_{-70}$pc, \citealp{dachs89}; 407$^{+22}_{-20}$pc, \citealp{terndrup02}; 404$^{+7}_{-8}$ pc, \citealp{an07}) and has a metallicity close to solar (e.g. [Fe/H]=-0.10$\pm$0.04, \citealp{sung02}; [Fe/H]=+0.01$\pm$0.07, \citealp{terndrup02}). Moreover, with a mass that is approximately twice that of the Pleiades it is relatively rich \citep[$M_{\rm tot}$=1240-1560M$_{\odot}$, ][]{jeffries01b}. A deep UV and red photographic survey of 25 square degrees of sky centered on the cluster was performed by \citet[][]{reimers82} and led to the identification of nine blue candidate white dwarf members. Ultimately, four of these were confirmed through spectroscopy to be DA cluster members and found to have masses significantly greater than the objects populating the peak of the field white dwarf mass distribution \citep[NGC\,2516-1, -2, -3 and -5, ][]{reimers82, koester96}.  Although early optical spectra, obtained with the European Southern Observatory's (ESO) 3.6m telescope, were of sufficient quality to allow confirmation of the trend that higher mass stars evolve to become more massive degenerates and to argue that the maximum mass of a white dwarf progenitor is $M_{\rm init}$$\simgreat$6M$_{\odot}$ \citep[e.g.][]{koester96}, their rather modest signal-to-noise at $\lambda$$\simless$4000\AA, suggests significant further reduction of the statistical uncertainties on the mass and cooling time estimates for these objects could be achieved with modern data.

Therefore, we re-observed the white dwarf members of NGC\,2516 with ESO's 8m Very Large Telescope (VLT) UT2 and the FOcal Reducer and low dispersion Spectrograph (FORS1).  A detailed description of the FORS instruments may be found on the ESO webpages\footnote{http://www.eso.org/instruments/fors2/}. The data were acquired in service mode within the periods 2007/04/24-27 and 2007/10/06-11/21. Since these targets are comparatively bright, we specified fairly relaxed constraints on the sky conditions and thus these observations were generally undertaken in poorer seeing and/or with some cloud present. All the FORS1 data were acquired using the 2$\times$2 binning mode of the $E2V$ CCD, the 600B+12 grism and a 1.6" slit which gives a notional resolution of $\lambda$/$\Delta$$\lambda$$\sim$500. Flat and arc exposures were obtained within a few hours of the acquisition of each of the science frames. The spectra were reduced and extracted following the procedure outlined in Section~\ref{specreduc}. A spectrum of the featureless DC white dwarf WD0000+345 was obtained with an identical set-up during this period and was used to remove the remaining instrumental response. We have used a grid of synthetic spectra and followed the procedures described in Sections~\ref{ngc3532:specmod} and \ref{ngc3532:mass} to re-determine the effective temperatures and surface gravities of these four stars (Table~\ref{wdmass2}). To serve as a check on the repeatability of the output from our spectral analysis procedure, NGC\,2516-1 was re-observed by us with the VLT UT1 and FORS2 on 2010/02/06 (see Section{~\ref{specreduc}). The excellent agreement between the parameter estimates for NGC\,2516-1 we derive from the two distinct sets of observations, lends confidence to our reduction and analysis process. Subsequently, for internal and external consistency, we have used the CO core, thick H-layer evolutionary models of \citet{fontaine01} to determine their masses and their cooling times (Table~\ref{wdmass2}).

\begin{table}
\begin{minipage}{80mm}
\begin{center}
\caption{Inferred progenitor masses of all the known white dwarf members of NGC\,2516 for the most likely cluster age (column 2) and plausible limiting ages (columns 3 and 4). The white dwarf cooling times were derived using CO core tracks and the stellar lifetimes have been transformed to masses using the solar metallicity evolutionary models of \citet{girardi00}.}
\begin{tabular}{lccc}
\hline
\multicolumn{1}{c}{ID} & \multicolumn{3}{c}{$M_{\rm init}$(M$_{\odot}$)}\\

\hline
Adopted cluster age:  &  150Myr & 120Myr & 180Myr \\ 
\hline
NGC\,2516-1 & 5.35$^{+0.31}_{-0.22}$ & 6.26$^{+0.58}_{-0.43}$   & 4.84$^{+0.20}_{-0.16}$ \\\\
NGC\,2516-2 & 4.91$^{+0.14}_{-0.11}$ & 5.45$^{+0.24}_{-0.16}$   & 4.51$^{+0.11}_{-0.09}$ \\\\
NGC\,2516-3 & 5.50$^{+0.39}_{-0.26}$ & 6.56$^{+0.67}_{-0.52}$   & 4.94$^{+0.22}_{-0.18}$ \\\\
NGC\,2516-5 & 5.19$^{+0.23}_{-0.17}$ & 5.94$^{+0.46}_{-0.32}$   & 4.73$^{+0.16}_{-0.14}$ \\

\hline
\label{progmassngc2516}
\end{tabular}
\end{center}
\end{minipage}
\end{table}

From the earlier ESO observations, \cite{koester96} determined that NGC\,2516-2 and NGC\,2516-5 most likely have $M$$>$1.0M$_{\odot}$, with a best estimate for the former of $M$=1.05M$_{\odot}$. \cite{claver01} used these results and the stellar evolutionary models of \citep{girardi00} to infer the masses of their progenitor stars to be $M_{\rm init}$$\sim$6M$_{\odot}$. Since these latter estimates are consistent with the initial masses at and above which stars of solar metallicity are predicted by that generation of the Padova group calculations to reach sufficiently high central temperatures to ignite C and, also, the former values are larger than the H-depleted core masses in these models ($M$=0.97M$_{\odot}$), it has been suggested that the NGC\,2516 white dwarfs may be composed of O and Ne \citep[e.g.][]{weidemann05}.
However, based on our new, improved, spectroscopy we find that it is most likely that all four (known) white dwarf members of the cluster have $M$$<$1.0M$_{\odot}$. This places them below more recent theoretical predictions for the minimum mass of ONe core white dwarfs \citep[e.g. $M$$\sim$1.05-1.1M$_{\odot}$, ][]{siess07, gilpons03}. All four of these objects also now appear to lie within a fairly narrow (final) mass range (Table~\ref{wdmass2}) despite extending over $\sim$30Myr in the cooling age domain (corresponding to $\sim$0.5M$_{\odot}$ variation in initial mass e.g. Table~\ref{progmassngc2516}). Some stellar evolutionary models predict that the masses of ONe remnants are a strong function of initial mass \citep[e.g.][]{iben85}. Moreover, at the most probable age for this cluster, the inferred progenitor masses are now all some way below 6M$_{\odot}$. So, in light of these new observational results and more recent theoretical predictions we propose that the cores of the NGC\,2516 white dwarfs are composed of C and O rather than O and Ne. If there are ONe core UMWDs within this cluster, they remain to be found. 

As the presence of a close companion can substantially impact the evolution of a star, it is also worth briefly discussing here the new near-IR imaging ($J$ and $K_{\rm S}$) of these objects recently obtained as part of the ESO Visible and Infrared Survey Telescope for Astronomy \citep[VISTA, ][]{emerson06,dalton06} Hemisphere Survey. This reveals no evidence of unresolved cool companions with $M$$>$0.04M$_{\odot}$ to NGC\,2516-1, -3 or -5. However, this data hints at a possible near-IR excess to NGC\,2516-2 (see Figure~\ref{irjband} and Table~\ref{wdmass2}). These observations were obtained during December 2009 in modest seeing ($\sim$1$^{\prime\prime}$). Assuming the source detected at 07$^{\rm h}$57$^{\rm m}$50.79$^{\rm s}$ -60$^{\circ}$49$^{\prime}$55.1$^{\prime\prime}$ (J2000.0) in the $J$ band imaging is associated with NGC\,2516-2 (there is no detection in the shallower $K_{S}$ imaging, see Table~\ref{wdmass2}), then we appear to be observing about a factor two too much flux in this waveband (we estimate a $<$10\% likelihood of a chance alignment based on the density of sources of similar or greater brightness in the image). According to the grids of synthetic photometry of \cite{holberg06} a $T_{\rm eff}$=35000K, log $g$=8.5 white dwarf could be expected to have $V$-$J$=-0.71 (M$_{J}$=11.2) but from the VISTA photometry we estimate that NGC\,2516-2 has $V$-$J$=0.19$\pm$0.20. Although we have not applied any transformations to either the synthetic or the observed photometry to account for the slight differences in the filter systems, any discrepancies are expected to be very small. Moreover, foreground reddening towards the cluster is relatively low, E($V$-$J$)$\approx$0.16, based on E($B$-$V$)=0.07 \citep{kharchenko05a} and the interstellar extinction curve of \cite{fitzpatrick99}. We note that according to the NEXTGEN models of \cite{baraffe98}, an object with M$_{J}$$\sim$11 (M$_{K_{S}}$$\sim$10) has a mass of only $\sim$50M$_{\rm Jupiter}$ at the age of NGC\,2516. Deeper $J$ and $K$ data for NGC\,2516-2 should be obtained, in good seeing, to confirm or otherwise the presence of a substellar companion.

\begin{figure}
\begin{center}
\includegraphics[angle=270,width=6cm]{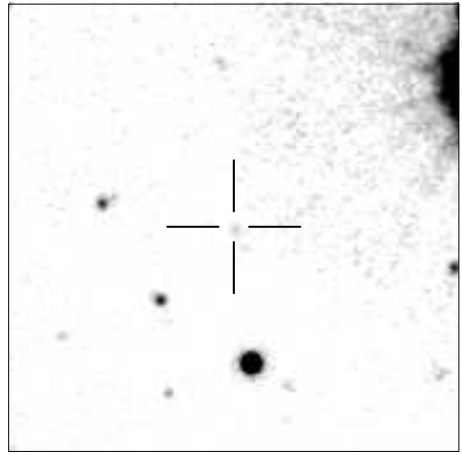}
\end{center}
\caption{A 1$^{\prime}$$\times$1$^{\prime}$ $J$ band image centered on NGC\,2516-2 (N top, E left). A weak source is seen within $\sim$1'' of the position of this white dwarf as measured from photographic plates \citep{reimers82}.}
\label{irjband}
\end{figure}

 \begin{figure*}
\includegraphics[angle=270,width=\linewidth]{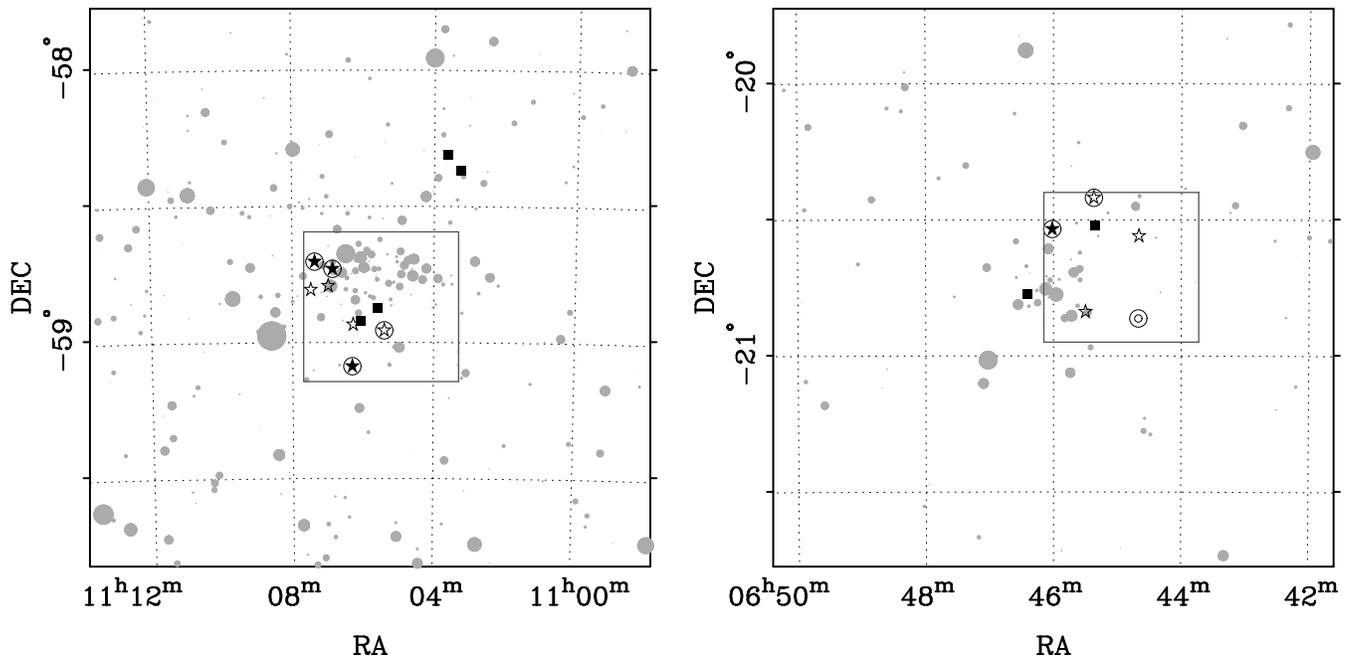}
\caption{Schematic plots of the open clusters NGC\,3532 (left) and NGC\,2287 (right) with the regions in each which have been surveyed with the ESO 2.2m + WFI, highlighted (large open rectangles). Candidate white dwarf members of the clusters identified from their location in $B-V$,$V$ colour-magnitude diagrams (five-point stars, both solid and open) and objects which have been followed-up spectroscopically with the VLT + FORS (open circles) are overplotted. The locations of previously known white dwarf members of the two clusters are also shown (squares). 
}
\label{SPATIAL}
\end{figure*}

%
%
\section{Observations and selection of targets}
\label{ngc3532:sample}
\subsection{The optical imaging surveys}
\label{ngc3532:sample_ESOWFI}
We retrieved $B$ and $V$ band imaging covering $\sim$0.3 square degrees of sky towards each of NGC\,2287 and NGC\,3532 from
the ESO data archive\footnote{http://archive.eso.org/cms/}(see Figure~\ref{SPATIAL}). These observations, which consist 
of a short (20 or 30s) and two 240s integrations per filter, were undertaken as part of the ESO Imaging Survey on the nights of 1999/12/03 (dark) and 
2000/02/24 (bright) respectively, with the Wide Field Imager \citep[WFI;][]{baade99} and the 2.2m telescope, located at La Silla, Chile. The ESO WFI consists 
of a mosaic of 
eight $EEV$4096$\times $2048 pixel CCDs and covers an area of 34'$\times$33' per pointing. The data were reduced using the Cambridge Astronomical Survey 
Unit CCD reduction toolkit \citep{irwin01}. We followed standard procedures, namely, subtraction of the bias, flat-fielding, astrometric calibration and 
co-addition. Subsequently, we performed aperture photometry on the reduced images using a circular window with a diameter of 1.5$\times$ the full width 
half maximum of the mean point spread function ($\sim$1.0-1.3''). Finally, we morphologically classified all sources detected in the combined frames and constructed merged catalogues of photometry.

\begin{figure*}
\includegraphics[angle=270,width=\linewidth]{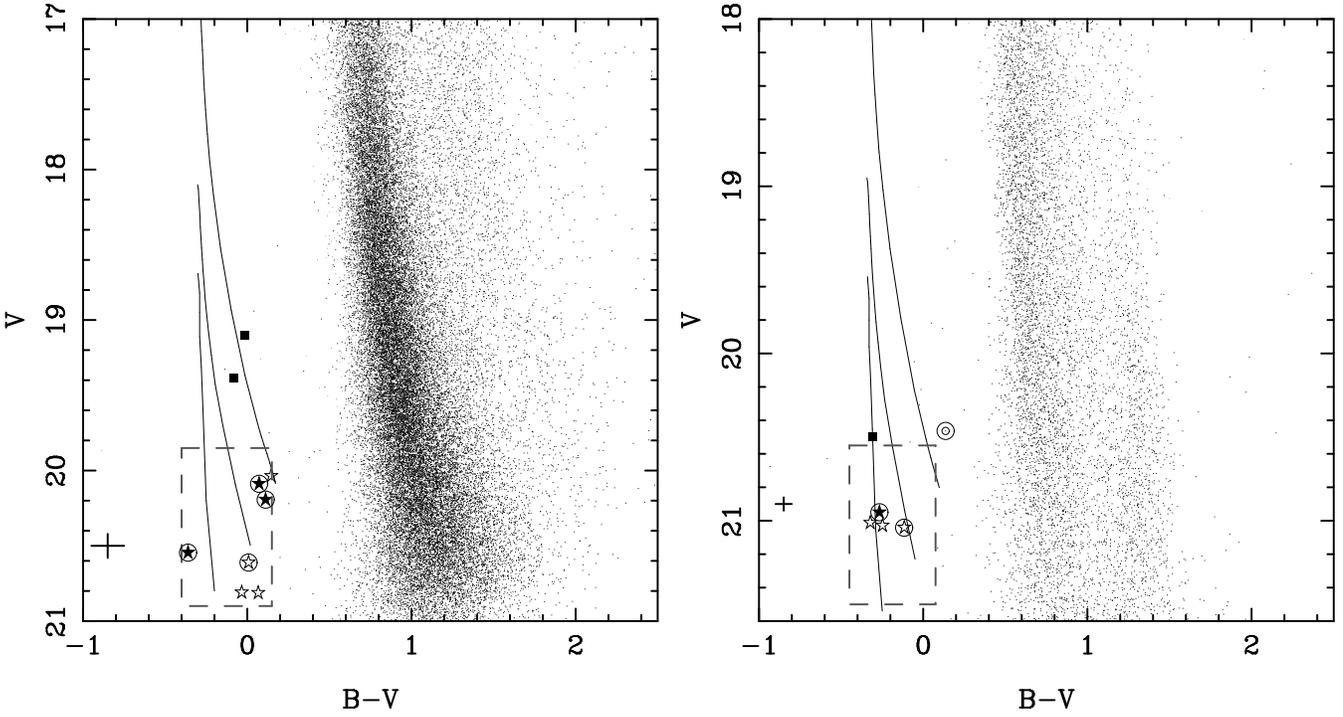}
\caption{$B-V$,$V$ colour-magnitude diagrams for all objects classified as stellar (dots) in the ESO WFI imaging of NGC\,3532 (left) and NGC\,2287 (right). As a guide 
to the likely location of the cluster white dwarfs we have highlighted the previously known cluster white dwarfs recovered here (squares) and have also overplotted evolutionary tracks for 0.7M$_{\odot}$, 1.1M$_{\odot}$ \citep[both CO core, thick H-layer; ][]{fontaine01} and 1.16M$_{\odot}$ \citep[ONe core, thick H-layer; ][]{althaus07} white dwarfs. These tracks have been shifted to distances of $V$-$M_{V}$=8.35 and $V$-$M_{V}$=9.30 and adjusted to allow for a foreground reddenings of $E$($B$-$V$)=0.05 and $E$($B$-$V$)=0.01  for NGC\,3532 and NGC\,2287 respectively. The remaining symbols to the objects are as defined in the caption of Figure~\ref{SPATIAL}.
}
\label{CMD}
\end{figure*}

As these archival datasets are more than a decade old, detailed sky transparency information (e.g. LOSSAM) for the time at which these observations were conducted is not available to us. Accordingly, to gauge the sky conditions we have examined the instrumental magnitudes of a series of bright (but not saturated) stars in the three sequential sub-integrations per filter. These were found to exhibit only small systematic differences of the order of a few percent from frame to frame. We have taken this to indicate that the sky was largely clear at the time of these observations. Subsequently, to convert the $B$ and $V$ instrumental magnitudes onto the standard Johnson system, we adopted zeropoints, colour terms and airmass co-efficients of k$_{0}$=24.53 and 24.12\footnote{www.eso.org/sci/facilities/lasilla/instruments/wfi/inst/zeropoints/ColorEquations/}, k$_{1}$=0.31 and -0.09 and k$_{2}$=0.20 and 0.11\footnote{www.eso.org/sci/facilities/lasilla/telescopes/d1p5/misc/Extinction.html}, respectively. While we have not performed a rigorous photometric calibration of these data, we believe the accuracy of our magnitudes, 
where sufficient photons have been detected, is good to $\sim$10\%. This is sufficient for the purpose of our initial surveys. We estimate from the number of point-like sources detected in the combined frames that these datsets are $\simgreat$50\% complete at $B$,$V$$\approx$21.5 and $B$,$V$$\approx$22 for NGC\,3532 and NGC\,2287 respectively.

\begin{table}
\begin{minipage}{80mm}
\begin{center}
\caption{A summary of the distances and reddenings we adopted in our photometric 
search for candidate white dwarf members of NGC2287 and NGC3532 and the 
cluster ages and metallicities we assumed in estimating progenitor masses.}
\begin{tabular}{lcccc}
\hline
Cluster  & $V$-$M_{V}$ & $E$($B$-$V$) & $\tau$ (Myr) & [Fe/H] \\ 
\hline
NGC\,2287 & 9.30  &  0.01   &  243$\pm$40 & 0.00 \\   
NGC\,3532 & 8.35  &  0.05   &  300$\pm$25 & 0.00 \\

\hline
\label{cluster_det}
\end{tabular}
\end{center}
\end{minipage}
\end{table}

\subsection[]{Selection of candidate white dwarf members of NGC\,3532 and NGC\,2287}

We have used our catalogues of photometry to construct $B$--$V$,$V$ colour magnitude diagrams for all point-like sources in the optical data 
(Figure~\ref{CMD}). As a guide to the location of the white dwarf cooling sequence of each cluster we have highlighted the previously known degenerate 
members within the surveyed areas (filled squares). Additionally, we have overplotted evolutionary tracks for $M$=0.7M$_{\odot}$, $M$=1.1M$_{\odot}$ \citep[CO
core, thick H-layer;][]{fontaine01} and $M$=1.16M$_{\odot}$ \citep[ONe core, thick H-layer;][]{althaus07} white dwarfs. These have been modified to account for
the distance to and foreground reddening towards each cluster i.e. $V$-$M_{V}$=8.35, $E$($B$-$V$)=0.05 for NGC\,3532 and $V$-$M_{V}$=9.30, $E$($B$-$V$)=0.01 for NGC\,2287 
(see Table~\ref{cluster_det}). Re-assuringly, a glance at Figure~\ref{CMD} reveals that the previously known white dwarf cluster members recovered here lie close to these theoretical tracks.

Since we are primarily interested in identifying the oldest, most massive cluster white dwarfs, for each population we have used the $V$ 
magnitude of the faintest previously known degenerate member ($V$$\approx$19.8 for NGC\,3532 and $V$$\approx$20.5 for NGC\,2287) as a guide
as to where to set the bright limits of our candidate selection boxes (dashed lines). Moreover, as the lower end of the theoretical tracks
plotted in Figure~\ref{CMD} correspond to the expected locations of white dwarfs which have cooled for a time approximately equal to the maximum likely age of each cluster ($\tau$$\approx$330Myr for NGC\,3532 and $\tau$$\approx$280Myr for NGC\,2287), we have used these to define the
faint limits of our selection boxes. In determining our selection criteria with respect to colour, we have adopted as the 
red limit the coolest point of the three evolutionary tracks plotted in Figure~\ref{CMD}. This represents a $M$=0.7M$_{\odot}$ white dwarf at the maximum probable age of each cluster. The blue limit is defined by the colour of the $M$=1.16M$_{\odot}$ track and includes an allowance for the 
estimated uncertainty in the $B$-$V$ colours ($\sim$0.15 mag.). The application of these selection criteria leads to the identification of 
seven and four candidate white dwarf members of NGC\,3532 and NGC\,2287, respectively. Summary details for these objects, including their co-ordinates and provisional magnitudes, are given in Table~\ref{candidates}. We note that NGC\,2287-WDC J0646-2032 has been previously identified as a potential white dwarf member of NGC\,2287 \citep{romanishin80}.

\begin{table*}
\begin{minipage}{151mm}
\begin{center}
\caption{A summary of the astrometric and photometric properties of the eleven candidate white dwarf members of NGC\,3532 and NGC\,2287 unearthed by our $B$,$V$survey imaging of 0.3 sq. degrees of sky towards each cluster. In column 2 we include the ID number of candidates towards NGC\,3532 as listed in Table 2 of {\protect \cite{clem11}}.}
\label{candidates}
\begin{tabular}{llcccccc}
\hline
\multicolumn{1}{c}{Survey} & \multicolumn{1}{c}{Other} & RA & DEC & $V$ & $B$-$V$ & VLT+FORS\\ 
\multicolumn{1}{c}{Designation}  & \multicolumn{1}{c}{IDs} & \multicolumn{2}{c}{J2000.0} & & & $?$ \\
 & & ($^{\rm h}$ $^{\rm m}$ $^{\rm s}$$_{.}$) & ($^{\circ}$ ' ``$_{.}$) \\
\hline

NGC\,3532-WDC\,J1105-5857  & WD\,J1105-585, 198279 &11 05 23.86 & -58 57 23.00 & 20.6 & 0.01 &Y\\
NGC\,3532-WDC\,J1106-5856  & 161574 &11 06 16.87 & -58 56 04.90 & 20.0 & 0.15&N\\
NGC\,3532-WDC\,J1106-5905  & WD\,J1106-590, 160608 & 11 06 18.35 & -59 05 17.33 & 20.1 & 0.07&Y\\
NGC\,3532-WDC\,J1106-5843  & WD\,J1106-584, 134356 & 11 06 51.65 & -58 43 49.03 & 20.2 & 0.11&Y\\
NGC\,3532-WDC\,J1106-5847 & 128702 &11 06 59.07 & -58 47 33.85 & 20.8 & -0.03&N\\
NGC\,3532-WDC\,J1107-5842 & WD\,J1107-584, 109695 &11 07 21.97 & -58 42 12.54 & 20.5 & -0.36&Y\\
NGC\,3532-WDC\,J1107-5848 & 104388 &11 07 28.74 & -58 48 24.25 & 20.8 & 0.07&N\\

\hline

NGC\,2287-WDC\,J0644-2033 &          -& 06 44 39.84 & -20 33 34.08 &  21.0 & -0.25 &N\\
NGC\,2287-WDC\,J0645-2025 & WD\,J0645-202      & 06 45 22.07 & -20 25 09.67 &  21.0 & -0.12 &Y\\
NGC\,2287-WDC\,J0645-2050 &          -& 06 45 29.93 & -20 50 18.45 &  21.0 & -0.32 &N\\
NGC\,2287-WDC\,J0646-2032 & WD\,J0646-203, NGC\,2287-4 & 06 46 01.08 & -20 32 03.42 &  20.9 & -0.27 &Y\\

\hline
\end{tabular}
\end{center}
\end{minipage}
\end{table*}

\begin{figure}
\includegraphics[angle=0,width=\linewidth]{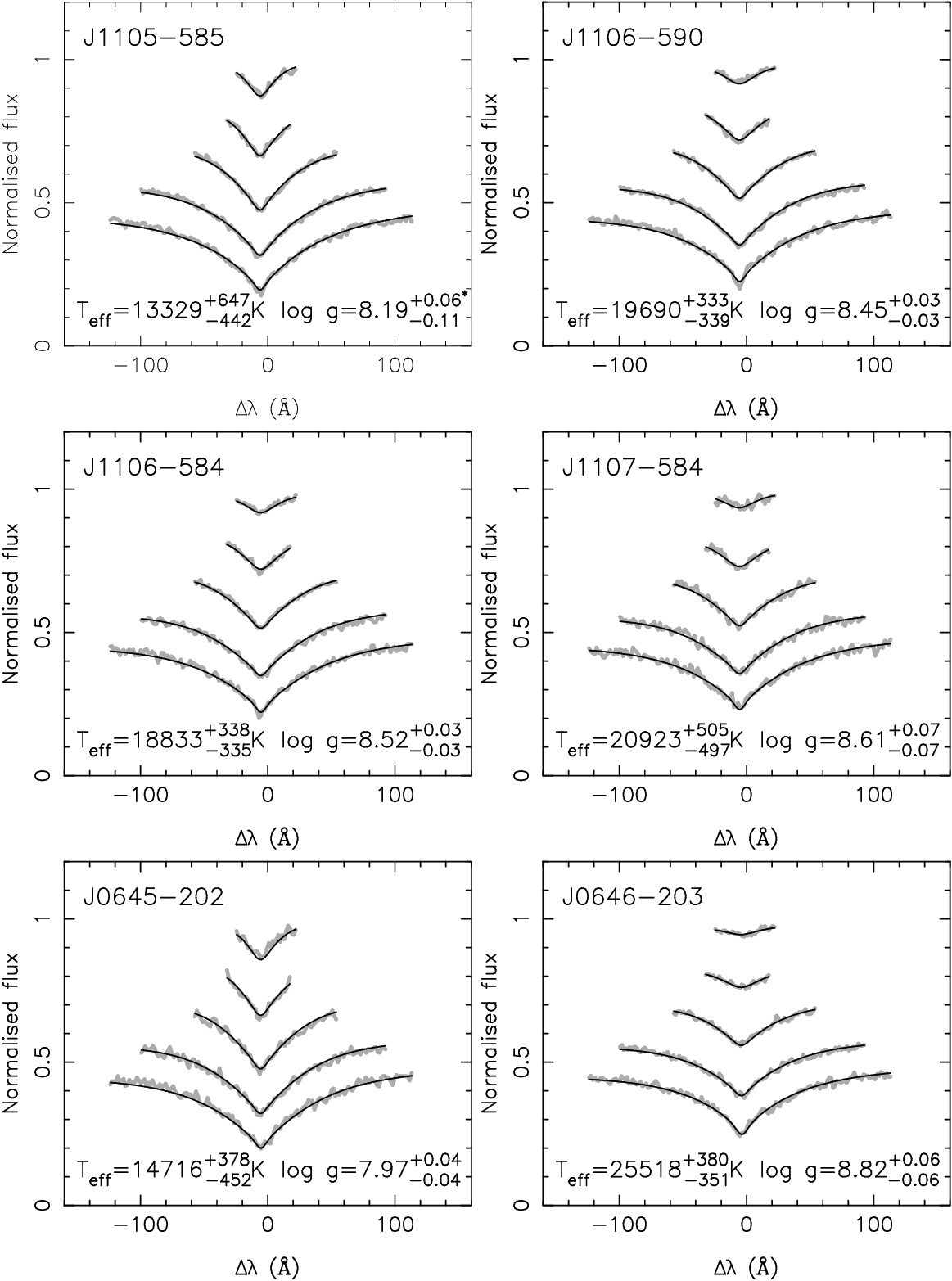}
\caption{The results of our fitting of synthetic profiles (thin black lines) to the observed Balmer lines, H-$\beta$ to H-8
, of the six white dwarf candidate members of NGC\,3532 and NGC\,2287 (thick grey lines). The flux$_{\lambda}$ units are arbitrary.}
   \label{LINES}
\end{figure}

%
%
%
%
\subsection{Follow-up spectroscopy with the VLT and FORS.}
\label{specreduc}

We have obtained follow-up low resolution optical spectroscopy spanning $\lambda$$\approx$3750-5200\AA\  for six of the candidates listed in Table~\ref{candidates} using the ESO VLT (UT1) and FORS2. These data were acquired in visitor mode on the nights of 2010/02/06-07. While seeing was generally good throughout this period, sky transparency on the first night was less than optimal with long spells of moderately thick cirrus. Fortunately, during the second night the sky was largely clear. The data were captured using the 2$\times$2 binning mode of the blue optimised $E2V$ CCD, the 600B+24 grism and a 1.3" slit, which provides a notional spectral resolution of $\lambda$/$\Delta$$\lambda$$\sim$600. Flat and arc exposures were obtained within a few hours of the acquisition of the science frames.

The CCD data were debiased and flat fielded using the {\tt{IRAF}} procedure {\tt{CCDPROC}}. Cosmic ray hits were removed using the routine 
{\tt{LACOS SPEC}} \citep{vandokkum01}. Subsequently the datasets were extracted using the {\tt{APEXTRACT}} package and wavelength calibrated
by comparison with He+HgCd arc spectra. Remaining instrumental signature was removed using a spectrum of the featureless DC 
white dwarf LHS\,2333 which was obtained using an identical set-up towards the beginning of our programme. The approximate S/N of the co-added datasets 
is listed in Table~\ref{wdmass}.

In error, we also spectroscopically observed a source in the direction of NGC\,2287 which our photometry indicates is not associated with this cluster (concentric open circles in Figures~\ref{SPATIAL}, right and~\ref{CMD}, right). We include this object in the remainder of our investigation as a test of our other membership selection criteria.

\section{Fundamental parameters of the white dwarf candidate members of NGC\,3532 and NGC\,2287}
\label{ngc3532:WDFP}

The energy distributions of all spectroscopically confirmed white dwarfs display the strongly pressure broadened H-Balmer 
lines characteristic of the spectral type DA. To obtain estimates of the fundamental parameters of our six candidates (and 
our test object) so that we can further scrutinise their cluster membership status, we have compared the Balmer lines in each
spectrum to synthetic line profiles obtained from model atmosphere calculations.

\subsection[]{The model atmosphere calculations}

\label{ngc3532:specmod}

The grid of synthetic pure-H spectra we use here is virtually identical to the one used in our previous work on these 
open clusters \citep[i.e.][]{dobbie09a} with the exception that it has been extended to slightly lower effective 
temperatures ($T$$_{\rm eff}$=12500-35000K). To re-iterate, our models have been generated using relatively recent versions
of the plane-parallel, hydrostatic, non-local thermodynamic equilibrium (non-LTE) atmosphere and spectral synthesis codes 
{\tt{TLUSTY}} (v200; Hubeny 1988, Hubeny \& Lanz 1995) and {\tt{SYNSPEC}} (v49; Hubeny, I. and Lanz, T. 2001, http://nova.astro.umd.edu/).
The model H atom employed in the calculations incorporates the 8 lowest energy levels and one superlevel extending from 
n=9 to n=80. Dissolution of the high lying levels has been treated by means of the occupation probability formalism of 
Hummer \& Mihalas (1988), generalised to the non-LTE situation by Hubeny, Hummer \& Lanz (1994). The calculations included
the bound-free and free-free opacities of the H$^{-}$ ion and incorporated a full treatment for the blanketing effects of HI
lines and the Lyman $-\alpha$, $-\beta$ and $-\gamma$ satellite opacities as computed by N. Allard (e.g. Allard et al. 2004). 
For models with $T$$_{\rm eff}$$>$16000K we assumed radiative equilibrium while at $T$$_{\rm eff}$$\le$16000K we 
included a treatment for convective energy transport according to the ML2 prescription of \cite{bergeron92}, with a 
mixing length parameter of $\alpha$=0.6. During the calculation of the model structure the hydrogen line broadening was addressed
in the following manner: the broadening by heavy perturbers (protons and hydrogen atoms) and electrons was treated using Allard's 
data (including the quasi-molecular opacity) and an approximate Stark profile (Hubeny, Hummer \& Lanz 1994) respectively. In 
the spectral synthesis step detailed profiles for the Balmer lines were calculated from the Stark broadening tables of Lemke (1997).

\begin{table*}
\begin{minipage}{172mm}
\begin{center}
\caption{Details of the new white dwarf candidate members of NGC\,3532 (upper) and NGC\,2287 (lower) and the ``test'' object in the direction 
of the latter cluster (bottom) observed with the VLT. The tabulated spectroscopic signal-to-noise estimates correspond to per resolution element over the 
range $\lambda$=4150-4300\AA. 
Masses and cooling times for each star have been estimated using the mixed CO core composition ``thick H-layer'' 
evolutionary calculations of the Montreal Group \citep{fontaine01}. The errors in absolute magnitudes, masses and 
cooling times shown here are derived by propagating more realistic uncertainties in effective temperature and 
surface gravity of 2.3\% and 0.07dex respectively.}

\label{wdmass}
\begin{tabular}{ccccccccccc}
\hline
 WD & S/N &$T$$_{\rm eff}^{*}$ & log $g$$^{*}$ &  $V$ & M$_{V}$ & M &  $\tau_{c}$ & $\mu_{\alpha}$cos $\delta$ & $\mu_{\delta}$  & Mem$?$  \\
   \multicolumn{2}{c}{} & (K)  &  \multicolumn{3}{c}{} & (M$_{\odot}$)  & (Myr) & (mas yr$^{-1}$) & (mas yr$^{-1}$)  &\\
\hline

 J1105-585 & 120 &$13329^{+647}_{-442}$ & $8.19^{+0.06^{*}}_{-0.11}$ & 20.33$\pm$0.04 & $11.75^{+0.11}_{-0.11}$ & $0.72\pm0.04$ & $370^{+50}_{-43}$  & +4.5$\pm$4.1 & -6.9$\pm$4.1 & $\times$ \\ \\
 J1106-590 & 115 &$19690^{+333}_{-339}$ & $8.45^{+0.03}_{-0.03}$ & 20.04$\pm$0.03  & $11.51^{+0.13}_{-0.13}$ & $0.90\pm0.04$ & $188^{+29}_{-26}$   & -7.9$\pm$3.4 & +0.5$\pm$3.4 & $\surd$ \\ \\
 J1106-584 & 100 &$18833^{+338}_{-335}$ & $8.52^{+0.03}_{-0.03}$ & 20.16$\pm$0.04 & $ 11.71^{+0.13}_{-0.13}$ & $0.94\pm0.04$ & $243^{+37}_{-32}$    & -5.7$\pm$3.8 & -1.0$\pm$3.6 & $\surd$ \\ \\
 J1107-584 & 80 &$20923^{+505}_{-497}$ & $8.61^{+0.07}_{-0.07}$ & 20.19$\pm$0.04 & $11.69^{+0.14}_{-0.13}$ & $1.00\pm0.04$ & $210^{+33}_{-29}$    & -3.9$\pm$3.9 & +4.6$\pm$4.3 & $\surd$ \\

\hline
 J0645-202 & 60 &$14716^{+378}_{-452}$ & $7.97^{+0.04}_{-0.04}$ & 20.97$\pm$0.04 & $11.26^{+0.11}_{-0.11}$ & $0.59\pm0.04$ & $200^{+30}_{-26}$   & +3.1$\pm$4.3 & +9.9$\pm$6.4 & $\times$ \\ \\
 J0646-203 & 115 &$25520^{+380}_{-351}$ & $8.82^{+0.06}_{-0.06}$ & 20.91$\pm$0.03 & $ 11.73^{+0.15}_{-0.15}$ & $1.12\pm0.04$ & $175^{+27}_{-25}$  & -4.2$\pm$3.4 & -0.3$\pm$2.6 & $\surd$ \\
\hline
 J0644-205 & 90 &$12376^{+230}_{-233}$ & $7.98^{+0.06}_{-0.05}$ & 20.52$\pm$0.03 & $11.57^{+0.10}_{-0.10}$ & $0.59\pm0.04$ & $334^{+41}_{-35}$  & +6.9$\pm$3.3 & -3.4$\pm$3.3 & $\times$ \\
\hline
\end{tabular}
\end{center}
$^{*}$ Formal fit errors - see text for further details. 
\end{minipage}
\end{table*}

\subsection{Effective temperatures and surface gravities}

\label{ngc3532:mass}

As in our previous work \citep[e.g.][]{dobbie09b,dobbie12a}, the comparison between the models and the data is undertaken
using the spectral fitting program {\tt{XSPEC}} \citep[][]{shafer91}. In the present analysis all lines from H-$\beta$ to H-8 are included
in the fitting process. {\tt{XSPEC}} works by folding a model through the instrument response before comparing the result to the data by means
of a $\chi^{2}-$statistic. The best fit model representation of the data is found by incrementing free grid parameters in small steps,
linearly interpolating between points in the grid, until the value of $\chi^{2}$ is minimised. Formal errors in the $T$$_{\rm eff}$s and 
log $g$s are calculated  by stepping the parameter in question away from its optimum value and redetermining minimum $\chi^{2}$ until the 
difference between this and the true minimum $\chi^{2}$ corresponds to $1\sigma$ for a given number of free model parameters \citep
[e.g.][]{lampton76}. The results of our fitting procedure are given in Table~\ref{wdmass} and are shown overplotted on the data in Figure~\ref{LINES}. 

It should be noted that the formal $1\sigma$ parameter errors quoted here undoubtedly underestimate the true uncertainties. In our 
subsequent analysis we have adopted what are considered more realistic levels of uncertainty of 2.3\% and 0.07dex in effective temperature and surface gravity, 
respectively \citep[e.g.][]{napiwotzki99}.

\begin{table}
\begin{minipage}{80mm}
\begin{center}
\caption{The co-efficients (zero-point, airmass and colour term) determined for Equation 1, the transformation between instrumental magnitudes and Johnson $V$.}

\label{wdvag}
\begin{tabular}{ccc}
\hline
$K_{0}$ & $K_{1}$ & $K_{2}$ \\
\hline
27.447$\pm$0.020 & -0.142$\pm$0.016 & 0.0004$\pm$0.0062\\

\hline
\end{tabular}
\end{center}
\end{minipage}
\end{table}~

\subsection{Distances}

To place stringent constraints on the distances and to measure the proper motions of the seven new spectroscopically confirmed DA white dwarfs we have obtained additional observations with the Inamori Magellan Areal Camera and Spectrograph (IMACS) and the 6.5m Magellan Baade telescope. IMACS is a wide field imaging system (or multiobject spectrograph) in which light from the Gregorian secondary of the telescope is fed to one of two cameras, either short (f/2.5) or long (f/4.3). The former camera, which was used for these observations, contains a mosaic of eight 2048$\times$4096 pixel $E2V$ CCDs. It covers a 27.5'$\times$27.5' (0.20'' pixels) area of sky per pointing but suffers from strong vignetting at distances  $>$15' from the field center. 

\begin{figure}
\label{distplot}
\includegraphics[angle=270,width=\linewidth]{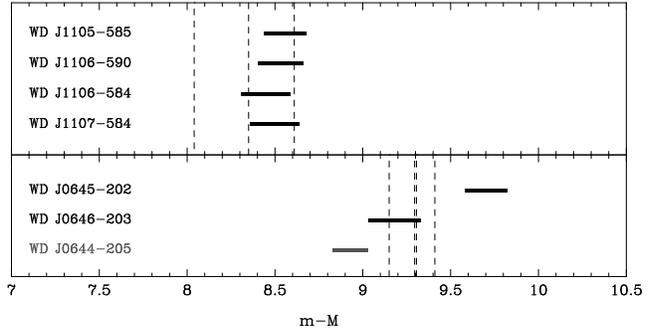}
\caption{The derived distance modulii of the white dwarf candidate members of NGC\,3532 (upper panel) and NGC\,2287 (lower 
panel) observed with the VLT and FORS. The estimated distance modulus of a control object lying in the field of NGC\,2287 is 
also shown (grey typeface). The distance modulus of NGC\,3532 as estimated by Meynet, Mermilliod \& Maeder (1993; $V$-$M_{V}$=8.35), Robichon et al. (1999; $m$-$M$=8.04) and Kharchenko et al. (2005; $V$-$M_{V}$=8.61) and the distance modulus of NGC\,2287 as estimated by Harris (1993; $V$-$M_{V}$=9.41), Meynet, Mermilliod \& Maeder (1993; $V$-$M_{V}$=9.15), Kharchenko et al. (2005; $V$-$M_{V}$=9.30) and Sharma et al. (2006; $V$-$M_{V}$=9.30) are overplotted. All new white dwarf candidate members have distances consistent with their proposed host clusters, except WD\,J0645-202 which appears to lie someway behind NGC\,2287.}
\end{figure}

\begin{figure*}
   \includegraphics[angle=270, width=\linewidth]{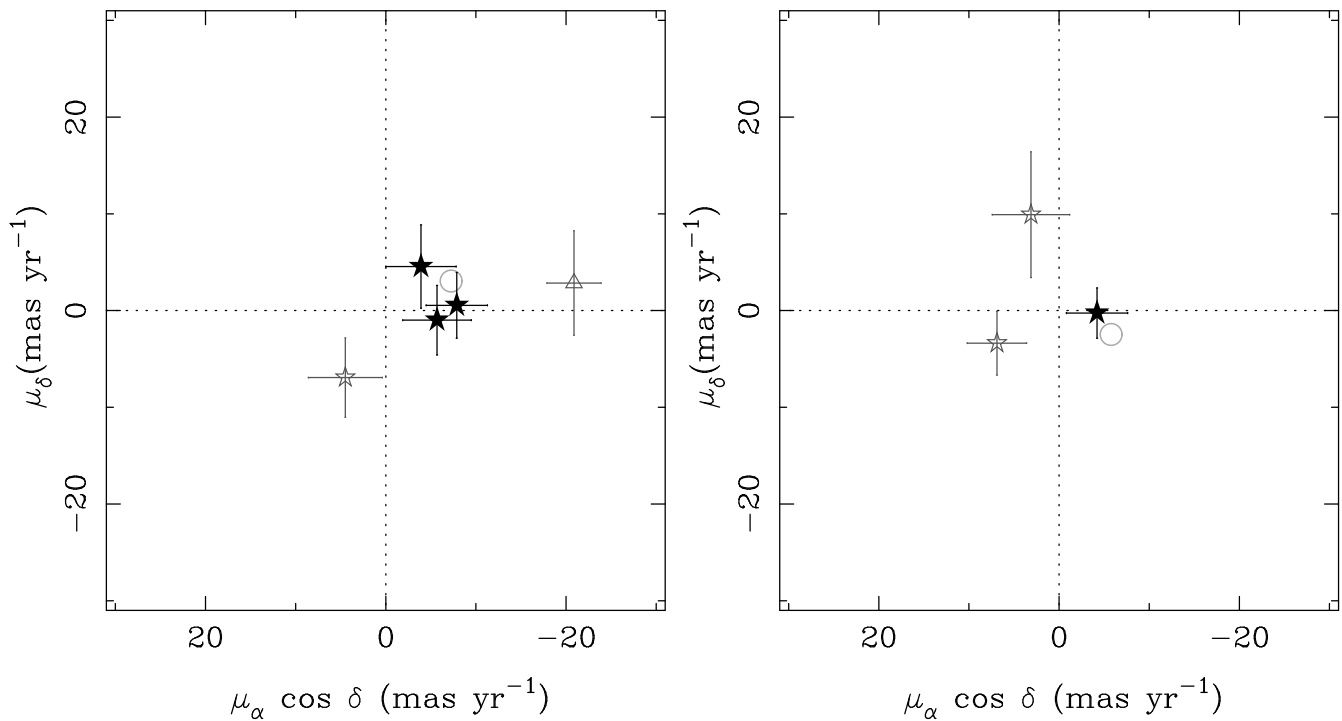}
   \caption{A vector point diagram of the relative proper motions of the seven spectroscopically observed white dwarfs towards NGC\,3532 (left) and NGC\,2287 (right).  The proper motions expected to be displayed by members of the two clusters are overplotted (grey open circles). These have been shifted to account for the difference between the relative and absolute frames of reference. Objects with proper motions deemed consistent with the cluster populations are highlighted (solid symbols). The proper motion of an additional photometric candidate white dwarf member of NGC\,3532 is also shown (open triangle, see text for further details).}
   \label{VPD}
\end{figure*}

We acquired 30\,s  integrations on each object through the $V_{\rm Bessell}$ filter during the photometric night of 2010/04/09 (JD=2455296.25) when the seeing was $\sim$0.7''. All targets were centered on CCD\,2 to negate the effects of chip-to-chip colour sensitivity variations. The frames were reduced, as per the ESO WFI data, using the Cambridge Astronomical Survey Unit CCD reduction toolkit \citep{irwin01} to follow the steps listed in Section~\ref{ngc3532:sample_ESOWFI}. As before, we performed aperture photometry on the reduced images using a circular window with a diameter of 1.5$\times$ the full width half maximum of the mean point spread function ($\sim$0.7'').  Several different standard star fields were also observed on CCD\,2 through both the $V_{\rm Bessell}$ and the $I_{\rm CTIO}$ filters during the course of this run to allow us to transform our instrumental magnitudes onto the Johnson system,

\begin{eqnarray}
m= -2.5 \log (ADU/t_{exp}) + K_{0} + K_{1}X + K_{2}(V-I)
\end{eqnarray}

\noindent where $ADU$ is a measure of the total counts from the source, $t_{exp}$ the exposure time and $X$ the airmass. The co-efficients and their respective errors were determined to have the values shown in Table~\ref{wdvag} and our revised $V$ magnitude estimates for the white dwarfs are given in Table~\ref{wdmass}. For the objects in the field of NGC\,2287, this more precise photometry is consistent with that based on the ESO WFI data. For the NGC\,3532 white dwarfs, our refined magnitudes are in excellent agreement with the values determined by \cite{clem11} but typically brighter than our preliminary WFI measurements by 0.1-0.3 mag. This suggests that our survey of this cluster is shallower by 0.1-0.3 mag. at $V$ than was initially concluded. 

We have used our measurements of the effective temperatures and surface gravities of the white dwarfs shown in Table~\ref{wdmass} 
and the model grids of \cite{bergeron95}, as revised by \cite{holberg06}, to derive absolute magnitudes (M$_{V}$; see Table~\ref{wdmass}).
Subsequently, we have determined the distance modulii of all seven white dwarfs observed with the VLT and FORS, neglecting foreground 
extinction which is low (A$_{V}$$\simless$0.15) along the lines of sight to NGC\,3532 and NGC\,2287. These are plotted 
(solid bars) along with a number of distance estimates available in the literature for each of these clusters (dash-dotted vertical lines) in
Figure~\ref{distplot}. Based on examination of this figure we consider that five out of the six white dwarf candidate cluster members have estimated
distances which are consistent with those of their putative parent populations (ie. within 2$\sigma$ of the mean). However, we find WD\,J0645-202 to lie 
substantially behind  NGC\,2287 and on this basis conclude that it is probably a field object. Not unexpectedly, the cooler, redder, ``test'' white dwarf, 
WD\,J0644-205, appears to reside in front of this cluster, albeit marginally.

\subsection{Proper motions}

To further explore the cluster membership credentials of these white dwarfs, we have exploited the epoch difference of approximately a 
decade between the archival ESO WFI $V$ band datasets and our new IMACS $V$ band images to measure their relative proper motions. Astrometry
is a potentially powerful method of discriminanting members of a cluster from the general field population \citep[e.g. ][]{hambly93}. We have
re-examined the images using the CASU reduction toolkit to determine the positions, in pixels, of bright (but not saturated), unblended, 
stellar-like objects lying within $\sim$2 arcminutes and on the same CCD chip as each candidate. We cross matched these lists of reference star 
positions using the {\tt{STARLINK TOPCAT}} software. Subsequently, we employed routines in the {\tt{STARLINK SLALIB}} library to construct six
co-efficient linear transforms between the two images of each candidate, where $\ge$3$\sigma$ outliers were iteratively clipped from the fits. The 
proper motions, in pixels, were determined by taking the differences between the observed and predicted locations of candidates in the 2nd epoch imaging.
These were then converted into milli-arcseconds per year in right ascension and declination using the world co-ordinate systems of the
datasets and dividing by the time baseline between the two observations ($\sim$10.12\,yr for NGC\,3532 and 10.35\,yr for NGC\,2287). The uncertainties 
on these measurements were estimated from the dispersion observed in the (assumed zero) proper motions of stars with similar brightness 
($\mid$$\Delta$$V$$\mid$$\simless$0.5 mag.) surrounding each white dwarf. The relative proper motion vector point diagrams for the objects towards each cluster 
are shown in Figure \ref{VPD} (five-point stars) and the measurements are listed in Table~\ref{wdmass}. The absolute proper motions of each cluster, on the Hipparcos system \citep{kharchenko05a,vanleeuwen09}, 
have been overplotted (grey open circles) after shifting them by $\mu$$_{\alpha}$ cos $\delta$, $\mu$$_{\delta}$ = +2.8, -1.7 mas yr$^{-1}$ and -1.2, -1.8 
mas yr$^{-1}$ for NGC\,3532 and NGC\,2287, respectively, to account for the difference between the relative and the absolute frames of reference. These 
offsets were estimated by taking the median of the absolute proper motions of stars at the faint end of the UCAC3 catalogue which reside within the survey areas. We note that there is a substantial overlap with our reference stars.

Of the three white dwarfs in the direction of NGC\,2287 that were observed with the VLT and FORS2, we find only WD\,J0646-203 (solid star in right hand panel) has a proper 
motion which we deem to be consistent with this cluster (ie. within 2$\sigma$). While three objects (solid stars in left hand panel) towards NGC\,3532 have proper motions 
within 1$\sigma$ of the cluster mean value, WD\,J1105-585 (left most open star) lies more than 3$\sigma$ away and on these grounds seems unlikely to be associated with 
NGC\,3532. The proper motion of NGC\,3532-WDC\,J1105-5842 (open triangle), an additional object which we were able to include in our astrometric study due to its spatial 
proximity to WD\,J1105-585, is found to be discrepant from that of the cluster at $>$4$\sigma$ ($\mu_{\alpha}$cos $\delta$, $\mu$$_{\delta}$=-20.9$\pm$3.0,+2.8$\pm$5.4 mas~yr$^{-1}$). This object has similar colours to WD\,J1106-590 and WD\,J1106-584, 
$B$-$V$$\approx$0.1 and despite lying close enough to the theoretical tracks in the $B$-$V$, $V$ CMD to be considered a photometric cluster member, was too bright ($V$$\approx$19.7)
to have featured in our spectroscopic follow-up program. 


%

\begin{table}
\begin{minipage}{80mm}
\begin{center}
\caption{Inferred progenitor masses of all the known white dwarf members of NGC\,2287 for the most likely cluster age (column 2) and plausible limiting ages (columns 3 and 4). The cooling times of all white dwarfs were initially derived using CO core tracks. The cooling time of WD\,J0646-203 was also evaluated using ONe models (bottom row). The estimated stellar lifetimes have been transformed to masses using the solar metallicity evolutionary models of \citet{girardi00}.}
\begin{tabular}{lccc}
\hline
\multicolumn{1}{c}{ID} & \multicolumn{3}{c}{$M_{\rm init}$(M$_{\odot}$)}\\
\hline
Adopted cluster age:  &  243Myr & 203Myr & 283Myr \\ 
\hline

NGC\,2287-2   & 4.45$^{+0.20}_{-0.15}$ & 4.99$^{+0.28}_{-0.22}$   & 4.08$^{+0.14}_{-0.10}$ \\\\
NGC\,2287-5   & 4.57$^{+0.24}_{-0.18}$ & 5.16$^{+0.37}_{-0.25}$   & 4.17$^{+0.16}_{-0.12}$ \\\\
WD\,J0646-203 & 6.41$^{+1.54}_{-0.87}$ & 9.73$^{*}_{-2.58}$      & 5.22$^{+0.68}_{-0.39}$ \\ \hline
WD\,J0646-203 & 5.94$^{+1.19}_{-0.60}$ & 8.06$^{+7.14}_{-1.34}$   & 5.02$^{+0.51}_{-0.32}$ \\

\hline
\label{progmassngc2287}
\end{tabular}
\end{center}
\end{minipage}
\end{table}

\section{New southern open cluster white dwarfs}

The four objects identified in Section~\ref{ngc3532:WDFP} as having properties consistent with cluster membership represent the faintest white dwarfs in NGC\,3532 and
 NGC\,2287 unearthed to date. Once again we have used the evolutionary models of \cite{fontaine01} to determine their masses and cooling times. For completeness 
we have also derived the masses and cooling times for the three white dwarfs in our spectroscopic sample which we have already rejected as probable field objects. So that this work 
is consistent with our previous efforts and those of several other recent studies in this area \citep[e.g][]{casewell09,dobbie09a,williams09,kalirai08} we have (initially) adopted the calculations which include a mixed CO core and thick H surface layer structure. 

\subsection{Is WD\,J0646-203 in NGC\,2287 a ONe core white dwarf$?$}

Based on our measurements of effective temperature and surface gravity we have determined the mass of WD\,J0646-203 to be $M$=1.12$\pm$0.04M$_{\odot}$ (Table~\ref{wdmass}). While the astrometric characteristics, the distance modulus and the projected location of this white dwarf only $\sim$0.37$^{\circ}$  from the cluster center (ie. within the projected tidal radius e.g. \citealt{cox54,piskunov08}) are firmly suggestive of an association with NGC\,2287, we have nonetheless explored the likelihood that it is merely a field star. At $T_{\rm eff}$$>$25000K, WD\,J0646-203 is hot enough to radiate significantly in the extreme-ultraviolet (EUV) and, had it been located within the Local Bubble, would likely have been detected in the EUV all-sky surveys \citep{pye95, bowyer96}. \cite{vennes97} have computed the space density of EUV detected white dwarfs to be n$\approx$1.6-2.2$\times$10$^{-5}$pc$^{-3}$. \cite{liebert05a} have estimated that only $\sim$1/5th of white dwarfs in the Palomar-Green survey have $M$$>$0.8M$_{\odot}$ while \cite{vennes97} find that 10 out of their sample of 90 have $M$$\simgreat$1.1M$_{\odot}$.  We estimate that as a result of flagging as potential members those objects with distance modulii consistent with the cluster mean at the 2$\sigma$ level, our survey has probed a volume of $\sim$9500pc$^{3}$. Thus, with the above space densities we should have expected to unearth $\sim$0.04 hot, massive field white dwarfs in our study of NGC\,2287. Assuming a poisson distribution, we determine that the probability of us detecting one field object with the characteristics of WD\,J0646-203 is rather low, P$\simless$0.04, and we are led to conclude that this white dwarf is a bona-fide member of NGC\,2287. 

Some theoretical studies suggest that the surface H-layers of the most massive white dwarfs may be of significantly lower mass than the 10$^{-4}$M$_{\odot}$ adopted in the ``standard'' evolutionary models \citep[e.g.][]{garcia97}. Consequently, we have re-evaluated the parameters of WD\,J0646-203 using the thin H-layer models of the Montreal group which allow for a surface H mass of only 10$^{-10}$M$_{\odot}$. In this case we obtain a slightly lower mass and marginally shorter cooling time but the differences between the two determinations are less than 0.02M$_{\odot}$ and 3Myr, respectively. Since we know the effective temperature of WD\,J0646-203 from our spectroscopy, to perform something of a sanity check on these mass determinations, we have also estimated it by assuming, a priori, that this white dwarf lies at the distance of the cluster, calculating its radius and applying a theoretical mass-radius relation. If the assumption of cluster membership is safe, the systematic uncertainty in the mass estimate, which relates to our spectroscopic analysis method and our choice of atmosphere code for generating the synthetic H line profiles \citep[e.g.][]{liebert05a}, should be reduced here. Using the effective temperature shown in Table~\ref{wdmass}, the arithmetic mean of the four distance modulii listed in the caption of Figure~\ref{distplot} (m$_{V}$-M$_{V}$=9.29$\pm$0.09) and the mass-radius relation from the thick and thin H-layer CO core evolutionary models we determine WD\,J0646-203 to have $M$=1.09$\pm$0.03M$_{\odot}$ and $M$=1.07$\pm$0.03M$_{\odot}$, respectively. 

\begin{table}
\begin{minipage}{80mm}
\begin{center}
\caption{Inferred progenitor masses of all the known white dwarf members of NGC\,3532 for the most likely cluster age (column 2) and plausible limiting ages (columns 3 and 4). The white dwarf cooling times were derived using CO core tracks and the stellar lifetimes have been transformed to masses using the solar metallicity evolutionary models of \citet{girardi00}.}
\begin{tabular}{lccc}
\hline
\multicolumn{1}{c}{ID} & \multicolumn{3}{c}{$M_{\rm init}$(M$_{\odot}$)}\\
\hline
Adopted cluster age:    & 300Myr & 275Myr & 325Myr \\
\hline
NGC\,3532-1     & 3.83$^{+0.09}_{-0.07}$ & 3.99$^{+0.10}_{-0.09}$  & 3.69$^{+0.08}_{-0.06}$\\\\
NGC\,3532-5     & 3.71$^{+0.06}_{-0.05}$ & 3.85$^{+0.07}_{-0.06}$  & 3.59$^{+0.06}_{-0.04}$\\\\
NGC\,3532-9     & 3.57$^{+0.02}_{-0.01}$ & 3.69$^{+0.02}_{-0.01}$  & 3.47$^{+0.01}_{-0.01}$\\\\
NGC\,3532-10    & 4.58$^{+0.35}_{-0.24}$ & 4.92$^{+0.43}_{-0.31}$  & 4.30$^{+0.28}_{-0.19}$\\\\
WD\,J1106-590   & 5.14$^{+0.69}_{-0.39}$ & 5.70$^{+1.19}_{-0.57}$  & 4.76$^{+0.47}_{-0.32}$ \\\\
WD\,J1106-584   & 6.92$^{+4.92}_{-1.28}$ & 9.01$^{+9.14}_{-2.42}$  & 5.86$^{+1.75}_{-0.75}$\\\\
WD\,J1107-584   & 5.64$^{+1.33}_{-0.59}$ & 6.58$^{+2.60}_{-1.05}$  & 5.11$^{+0.79}_{-0.42}$\\
\hline

\label{progmassngc3532}
\end{tabular}
\end{center}
\end{minipage}
\end{table}

These values argue that WD\,J0646-203 is potentially the most massive white dwarf yet identified within an open cluster. For example, LB1497, the sole white dwarf located within the projected tidal radius of the Pleiades cluster, has $M$=1.05$\pm$0.02M$_{\odot}$, based on a weighted mean of the estimates in \cite{bergeron95}, \cite{dobbie06a} and \cite{vennes08}, which were derived using thick H-layer CO core evolutionary models. The masses of the four most massive white dwarfs recently unearthed in the slightly metal poor cluster NGC\,2168, which were also derived using thick H-layer CO core tracks, appear to be clumped around $M$=1.01-1.02M$_{\odot}$ \citep{williams09}. Indeed, our CO core based mass determinations for WD\,J0646-203 sit on or above recent theoretical predictions for the minimum mass of ONe core white dwarfs \citep[e.g. $M$$\sim$1.05-1.1M$_{\odot}$, ][]{siess07, gilpons03}. The most likely progenitor mass we infer of $M_{\rm init}$$\sim$6.3-6.4M$_{\odot}$ (thin or thick H-layer white dwarf models) is also consistent with the initial mass range where some solar metallicity stellar evolutionary calculations, which include overshooting from the convective regions of the star, predict central C ignition \citep[$M_{\rm init}$$>$6M$_{\odot}$ e.g.][]{eldridge04, bertelli09}. While the ignition of C in the core is anticipated to occur only at larger initial masses (by $\sim$1-2M$_{\odot}$) in stellar models that neglect to treat this phenomenon, a number of distinct observational based results provide compelling arguments in favour of significant convective overshooting \citep[e.g.][]{woo03,claret07,salaris09}. Moreover, despite our inferred progenitor mass ($M_{\rm init}$$\sim$5.2M$_{\odot}$) for a cluster age towards the upper end of the plausible range for NGC\,2287 (see Table~\ref{progmassngc2287}), probably being too low for a star of this composition to have experienced central C burning, in this eventuality, WD\,J0646-203 would appear as something of an outlier in a semi-emprical IFMR, sitting $\simgreat$0.1M$_{\odot}$ above all other white dwarfs from solar metalicity populations located in this part of the initial mass domain. For example, the well studied Sirius B \citep{barstow05, liebert05b} and the NGC\,3532 and NGC\,2516 objects studied by us here using a similar spectroscopic set-up and the same models and analysis techniques cluster around $M$$\sim$0.95-1.0M$_{\odot}$. Of course, this variation could be statistical in nature but earlier we noted that our estimates of mass obtained using two different approaches are in good agreement.

Bearing in mind that objects of ONe composition are expected to adhere to a slightly different mass-radius relation, we have re-evaluated the parameters of WD\,J0646-203 once again, this time using the ONe core evolutionary tracks of \cite{althaus07}. Based on our spectroscopic effective temperature and surface gravity measurements we estimate a mass and a cooling time of $M$=1.09$\pm$0.04M$_{\odot}$ and $\tau_{\rm cool}$=164$^{+26 }_{-22}$Myr, respectively. Following our other approach, where we assume WD\,J0646-203 lies at the distance of NGC\,2287, we determine $M$=1.05$\pm$0.03M$_{\odot}$. These revised parameters for our object remain in accord with modern theoretical estimates of the minimum mass of an ONe core UMWD at approximately solar metallicity \citep[][]{siess07} and, for the most likely cluster age, lead to an inferred progenitor mass consistent with the initial mass range where central C ignition is anticpated (see Table~\ref{progmassngc2287}). Clearly it not possible for us to definitively determine the composition of the core of WD\,J0646-203 but on the basis of our new observations and guided by modern theoretical calculations, we conclude that there is a distinct possibility that it consists of O and Ne, having evolved from a single, heavy-weight, intermediate mass progenitor star. 

\subsection{The NGC\,3532 white dwarfs}

\begin{figure*}
\begin{center}
\includegraphics[angle=270,width=14cm]{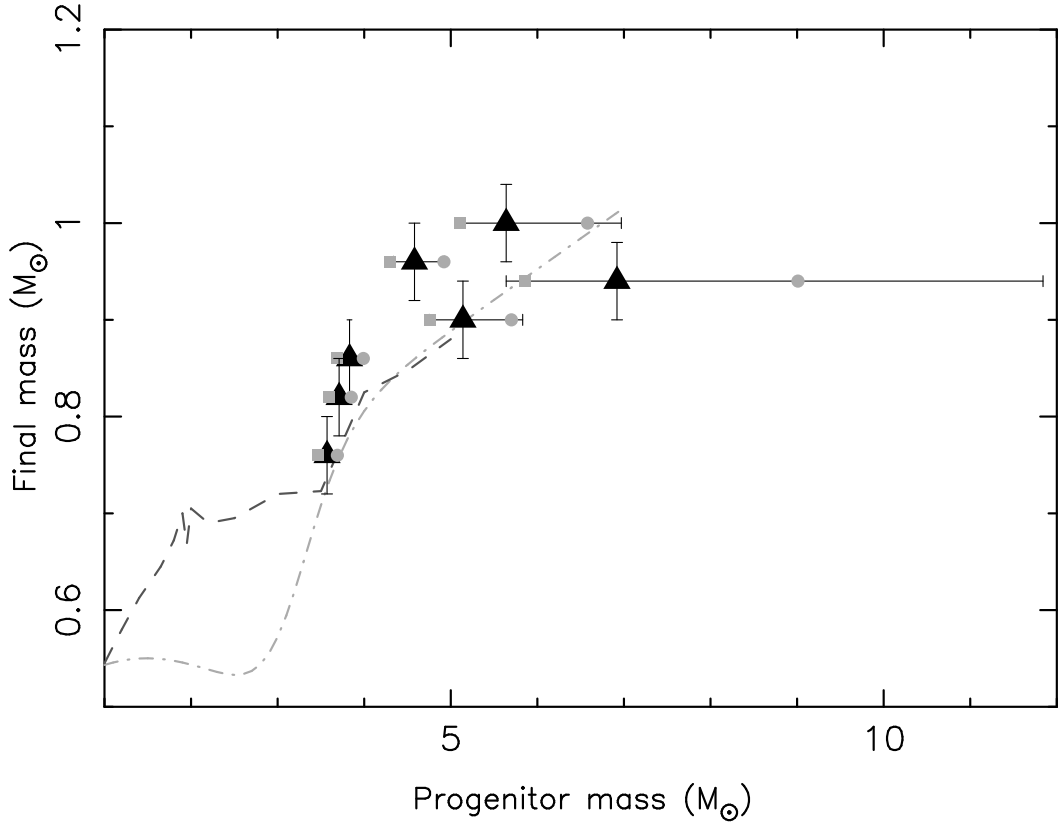}
\end{center}
\caption{The location of the seven known NGC\,3532 white dwarfs in initial mass-final mass space for the most likely cluster age (300Myr, black triangles) and approriate limiting ages (275Myr, grey circles and 325Myr, grey squares). A theoretical IFMR from \citet{marigo07} (dark-grey dashed line) and a 1TPR from \citet{wagenhuber98} (light-grey dot-dashed line) are overplotted.}
\label{ifmr}
\end{figure*}

In our previous work on NGC\,3532, we revisited six white dwarfs in the direction of the cluster and confirmed four as members \citep{dobbie09a}. Two objects, NGC\,3532-6 and NGC\,3532-8, were rejected by us on the basis that our revised estimates of their distances are incompatable with that of the cluster. We have now found three further white dwarfs members of NGC\,3532 taking the total number to seven and making this is one of the richest hauls of cluster degenerates where good quality optical spectroscopy has been secured. This population is important since NGC3532 has a turn-off mass of $M_{\rm init}$$\approx$3.5M$_{\odot}$ and spans the age gap between the younger NGC\,2168 and more mature Praesepe and Hyades clusters. 

Despite our new study probing to fainter magnitudes than the previous photographic survey of the cluster, we have not identified any compelling candidate ultra-massive ONe core white dwarf members of NGC\,3532 (ie. $M$$\simgreat$1.05M$_{\odot}$ and $M_{\rm init}$$\simgreat$6M$_{\odot}$). Instead, the three new fainter degenerate members have masses that are similar to that of the previously known faintest white dwarf in the cluster, NGC\,3532-10. Intruiguingly, despite their comparable masses, we find that the cooling times of these four most massive white dwarf members span a range of $\sim$90-100Myr, irrespective of our choice of thick or thin H-layer evolutionary models. For example, for WD\,J1106-584 we estimate $\tau_{\rm cool}$$\sim$240Myr while for NGC\,3532-10 we determined $\tau_{\rm cool}$$\sim$150Myr. We suggested in our previous study, where we analysed an ensemble of $\sim$50 white dwarfs from $\sim$10 mainly cluster populations, that the IFMR is less steep at $M_{\rm init}$$\approx$4M$_{\odot}$ than immediately below this initial mass regime. We noted that the distribution of these white dwarfs mirrored the form of the theoretical relationship between the initial mass of a star and the mass of the CO core at the time of the first thermal pulse (1TPR) as it evolves onto the asymptotic giant branch \citep[e.g.][]{becker79}. \cite{salaris09} independently reached a similar conclusion regarding the slope of the (semi-) empirical relation by comparing the heterogeneous open cluster data to an IFMR predicted by the {\tt BaSTI} stellar evolutionary models of \cite{pietrinferni04}. 

Nevertheless, given the rather heterogeneous nature of that ensemble of white dwarfs, it is conceivable that cluster to cluster age uncertainties could have conspired in such a way to mislead us. For example, if the ages of the youngest populations which dominate the upper end of the relation are systematically underestimated with respect to the more mature populations which dominate the lower end then the semi-empirical IFMR would appear less steep at higher initial masses. However, independent of both the age we adopt for NGC\,3532, within plausible limits, and our choice of white dwarf evolutionary model grid (thick or thin surface H-layers), the location in initial mass-final mass space of these seven degenerates (Figure~\ref{ifmr} and see also Table~\ref{progmassngc3532}), which span a broad range in initial mass yet are vulnerable to only a single cluster age uncertainty, are also consistent with the trend outlined by the 1TPR \citep[e.g.][]{wagenhuber98} and the theoretical IFMR \citep[e.g.][]{marigo07}. This lends some further support to both our earlier conclusion and that of \cite{salaris09} regarding the form of the IFMR. 

Population synthesis calculations which adopt an IFMR with a lower gradient above $M_{\rm init}$$\approx$4M$_{\odot}$, than in the immediately lower initial mass regime, better reproduce the shape of the field white dwarf mass distribution, in particular the secondary peak at $M$$\sim$0.8-1.0M$_{\odot}$, than those which employ a relation with a simple linear form \citep{ferrario05}. It has been proposed previously that  $\simgreat$80\% of white dwarfs with  $M$$>$0.8M$_{\odot}$ might originate from binary mergers \citep[e.g.][]{liebert05a} but such a change in the gradient of the IFMR mitigates the need to invoke such an evolutionary channel to explain their existence. That stellar evolution might be able to readily produce large CO cores without the need for mergers is potentially important in the context of Type Ia supernovae which are believed to result from the complete thermonuclear disruption of massive CO white dwarfs \citep[e.g][]{yungelson00}. For example, in a leading, current, theoretical formation channel, the sub-Chandrasekar double-detonation model, the more massive a CO core progenitor, the less material is required to be accreted to trigger the explosion. We note that the lifetimes of those stars to which the less steep section of the IFMR is applicable (ie. $M_{\rm init}$$\simgreat$4M$_{\odot}$) are comparable to the timescales of the prompt component of the Type Ia delay time distribution \citep[e.g. see][]{ruiter11}.

%
%
\section{Summary}

We have performed a CCD photometric survey, complimented by optical spectroscopic and astrometric follow-up studies, of $\sim$0.3 square degrees of sky towards each of the young/intermediate age open clusters NGC\,3532 and NGC\,2287, in a bid to unearth intrinsically faint white dwarf members. This work has led to the identification of four new white dwarfs which are probably cluster members, three in NGC\,3532 and one in NGC\,2287. Spectroscopic analysis of these objects has revealed WD\,J0646-203 to be potentially the most massive white dwarf member of an open cluster unearthed so far, with there being a distinct possibility of it having a core composed of O and Ne ($M$=1.02-1.16M$_{\odot}$). Our study has also shown that, despite a range of $\sim$90Myr in age, the four most massive degenerates in NGC\,3532 are rather similar in mass, consistent with our earlier suggestions that the IFMR could be less steep at $M$$_{\rm init}$$\simgreat$4M$_{\odot}$ than at initial masses immediately below this (Figure~\ref{ifmr}). The results of our analysis of new and improved spectroscopy of the four known NGC\,2516 white dwarfs are also in agreement with our prior conclusion. 

However, further studies of young/intermediate age, rich, open clusters remain necessary to better determine the relative frequencies of massive and ultra-massive degenerates and to pin down the maximum mass of a white dwarf that can form via canonical stellar evolution. It appears probable that additional constraints on the form of the IFMR will come from NGC\,3532 since a more extensive CCD survey of the cluster has recently unearthed many more faint candidate white dwarf members \citep{clem11}. Low resolution optical spectroscopy with good signal-to-noise is now required to characterise these objects. Substantial amounts of time on the current generation of large telescopes would be required to obtain spectroscopy of the faintest white dwarf members of attractive, even richer but more distant clusters. Alternatively, it might be possible to address this type of investigation photometrically since simulations suggest that in the $U$,$U$-$V$ colour-magnitude diagram of a very rich cluster with $\tau$$\sim$200-300Myr (e.g. NGC\,6705), a putative ultra-massive white dwarf population would define a rather prominent blue hook at the bottom end of the cooling sequence and give rise to a secondary peak in the $U$ band luminosity function. Detection of this feature would need photometry of good quality down to V$\approx$25 so even this approach could prove rather challenging given the likely levels of scattered light from the large number of intrinsically bright stars in rich clusters of this age.

\label{ngc3532:conclusions}
%
%
%
\section*{Acknowledgments}
Based on observations made with 
ESO Telescopes at the La Silla Paranal Observatory under programme 
numbers 164.O-0561, 80.D-0654 and 084.D-1097. Data from 164.O-0561  was 
retrieved from the ESO Science Archive Facility.
This research has made use of the Simbad database, operated at
the Centre de Donn\'ees Astronomiques de Strasbourg (CDS), and
of NASA's Astrophysics Data System Bibliographic Services (ADS). 
Based on observation obtained as part of the VISTA
Hemisphere Survey, ESO Progam, 179.A-2010 (PI: McMahon).
ADJ is supported by a FONDECYT fellowship, under project no. 3100098
NL acknowledges support from the national program number AYA2010-19136 funded 
by the Spanish Ministry of Science and Innovation.

%
%
\bibliographystyle{mn2e}
\bibliography{mnemonic,therefs}

\begin{thebibliography}{}

\bibitem[\protect\citeauthoryear{{Althaus}, {Garc{\'{\i}}a-Berro}, {Isern},
  {C{\'o}rsico} \& {Rohrmann}}{{Althaus} et~al.}{2007}]{althaus07}
{Althaus} L.~G.,  {Garc{\'{\i}}a-Berro} E.,  {Isern} J.,  {C{\'o}rsico} A.~H.,
    {Rohrmann} R.~D.,  2007, A\&A, 465, 249

\bibitem[\protect\citeauthoryear{{An}, {Terndrup} \& {Pinsonneault}}{{An}
  et~al.}{2007}]{an07}
{An} D.,  {Terndrup} D.~M.,    {Pinsonneault} M.~H.,  2007, ApJ, 671, 1640

\bibitem[\protect\citeauthoryear{{Baade}, {Meisenheimer}, {Iwert}, {Alonso},
  {Augusteijn}, {Beletic}, {Bellemann}, {Benesch} \& {30 co-authors}}{{Baade}
  et~al.}{1999}]{baade99}
{Baade} D.,  {Meisenheimer} K.,  {Iwert} O.,  {Alonso} J.,  {Augusteijn} T.,
  {Beletic} J.,  {Bellemann} H.,  {Benesch} W.,    {30 co-authors} 1999, The
  Messenger, 95, 15

\bibitem[\protect\citeauthoryear{{Baraffe}, {Chabrier}, {Allard} \&
  {Hauschildt}}{{Baraffe} et~al.}{1998}]{baraffe98}
{Baraffe} I.,  {Chabrier} G.,  {Allard} F.,    {Hauschildt} P.~H.,  1998, A\&A,
  337, 403

\bibitem[\protect\citeauthoryear{{Barstow}, {Bond}, {Holberg}, {Burleigh},
  {Hubeny} \& {Koester}}{{Barstow} et~al.}{2005}]{barstow05}
{Barstow} M.~A.,  {Bond} H.~E.,  {Holberg} J.~B.,  {Burleigh} M.~R.,  {Hubeny}
  I.,    {Koester} D.,  2005, MNRAS, 362, 1134

\bibitem[\protect\citeauthoryear{{Becker} \& {Iben} Jr.}{{Becker} \&
  {Iben}}{1979}]{becker79}
{Becker} S.~A.,  {Iben} Jr. I.,  1979, ApJ, 232, 831

\bibitem[\protect\citeauthoryear{{Bergeron}, {Saffer} \& {Liebert}}{{Bergeron}
  et~al.}{1992}]{bergeron92}
{Bergeron} P.,  {Saffer} R.~A.,    {Liebert} J.,  1992, ApJ, 394, 228

\bibitem[\protect\citeauthoryear{{Bergeron}, {Wesemael} \&
  {Beauchamp}}{{Bergeron} et~al.}{1995}]{bergeron95}
{Bergeron} P.,  {Wesemael} F.,    {Beauchamp} A.,  1995, PASP, 107, 1047

\bibitem[\protect\citeauthoryear{{Bertelli}, {Nasi}, {Girardi} \&
  {Marigo}}{{Bertelli} et~al.}{2009}]{bertelli09}
{Bertelli} G.,  {Nasi} E.,  {Girardi} L.,    {Marigo} P.,  2009, A\&A, 508, 355

\bibitem[\protect\citeauthoryear{{Bowyer}, {Lampton}, {Lewis}, {Wu}, {Jelinsky}
  \& {Malina}}{{Bowyer} et~al.}{1996}]{bowyer96}
{Bowyer} S.,  {Lampton} M.,  {Lewis} J.,  {Wu} X.,  {Jelinsky} P.,    {Malina}
  R.~F.,  1996, ApJS, 102, 129

\bibitem[\protect\citeauthoryear{{Burbidge}, {Burbidge}, {Fowler} \&
  {Hoyle}}{{Burbidge} et~al.}{1957}]{burbidge57}
{Burbidge} E.~M.,  {Burbidge} G.~R.,  {Fowler} W.~A.,    {Hoyle} F.,  1957,
  Reviews of Modern Physics, 29, 547

\bibitem[\protect\citeauthoryear{{Casewell}, {Dobbie}, {Napiwotzki},
  {Burleigh}, {Barstow} \& {Jameson}}{{Casewell} et~al.}{2009}]{casewell09}
{Casewell} S.~L.,  {Dobbie} P.~D.,  {Napiwotzki} R.,  {Burleigh} M.~R.,
  {Barstow} M.~A.,    {Jameson} R.~F.,  2009, MNRAS, 395, 1795

\bibitem[\protect\citeauthoryear{{Claret}}{{Claret}}{2007}]{claret07}
{Claret} A.,  2007, A\&A, 475, 1019

\bibitem[\protect\citeauthoryear{{Claver}, {Liebert}, {Bergeron} \&
  {Koester}}{{Claver} et~al.}{2001}]{claver01}
{Claver} C.~F.,  {Liebert} J.,  {Bergeron} P.,    {Koester} D.,  2001, ApJ,
  563, 987

\bibitem[\protect\citeauthoryear{{Clem}, {Landolt}, {Hoard} \&
  {Wachter}}{{Clem} et~al.}{2011}]{clem11}
{Clem} J.~L.,  {Landolt} A.~U.,  {Hoard} D.~W.,    {Wachter} S.,  2011, AJ,
  141, 115

\bibitem[\protect\citeauthoryear{{Cox}}{{Cox}}{1954}]{cox54}
{Cox} A.~N.,  1954, ApJ, 119, 188

\bibitem[\protect\citeauthoryear{{Dachs} \& {Kabus}}{{Dachs} \&
  {Kabus}}{1989}]{dachs89}
{Dachs} J.,  {Kabus} H.,  1989, A\&AS, 78, 25

\bibitem[\protect\citeauthoryear{{Dalton}, {Caldwell}, {Ward}, {Whalley},
  {Woodhouse}, {Edeson}, {Clark}, {Beard}, {Gallie}, {Todd}, {Strachan},
  {Bezawada}, {Sutherland} \& {Emerson}}{{Dalton} et~al.}{2006}]{dalton06}
{Dalton} G.~B.,  {Caldwell} M.,  {Ward} A.~K.,  {Whalley} M.~S.,  {Woodhouse}
  G.,  {Edeson} R.~L.,  {Clark} P.,  {Beard} S.~M.,  {Gallie} A.~M.,  {Todd}
  S.~P.,  {Strachan} J.~M.~D.,  {Bezawada} N.~N.,  {Sutherland} W.~J.,
  {Emerson} J.~P.,  2006, in Society of Photo-Optical Instrumentation Engineers
  (SPIE) Conference Series Vol.~6269 of Society of Photo-Optical
  Instrumentation Engineers (SPIE) Conference Series, {The VISTA infrared
  camera}

\bibitem[\protect\citeauthoryear{{Dobbie}, {Baxter}, {K{\"u}lebi}, {Parker},
  {Koester}, {Jordan}, {Lodieu} \& {Euchner}}{{Dobbie}
  et~al.}{2012}]{dobbie12a}
{Dobbie} P.~D.,  {Baxter} R.,  {K{\"u}lebi} B.,  {Parker} Q.~A.,  {Koester} D.,
   {Jordan} S.,  {Lodieu} N.,    {Euchner} F.,  2012, MNRAS, 421, 202

\bibitem[\protect\citeauthoryear{{Dobbie}, {Casewell}, {Burleigh} \&
  {Boyce}}{{Dobbie} et~al.}{2009}]{dobbie09b}
{Dobbie} P.~D.,  {Casewell} S.~L.,  {Burleigh} M.~R.,    {Boyce} D.~D.,  2009,
  MNRAS, 395, 1591

\bibitem[\protect\citeauthoryear{{Dobbie}, {Napiwotzki}, {Burleigh}, {Barstow},
  {Boyce}, {Casewell}, {Jameson}, {Hubeny} \& {Fontaine}}{{Dobbie}
  et~al.}{2006}]{dobbie06a}
{Dobbie} P.~D.,  {Napiwotzki} R.,  {Burleigh} M.~R.,  {Barstow} M.~A.,  {Boyce}
  D.~D.,  {Casewell} S.~L.,  {Jameson} R.~F.,  {Hubeny} I.,    {Fontaine} G.,
  2006, MNRAS, 369, 383

\bibitem[\protect\citeauthoryear{{Dobbie}, {Napiwotzki}, {Burleigh},
  {Williams}, {Sharp}, {Barstow}, {Casewell} \& {Hubeny}}{{Dobbie}
  et~al.}{2009}]{dobbie09a}
{Dobbie} P.~D.,  {Napiwotzki} R.,  {Burleigh} M.~R.,  {Williams} K.~A.,
  {Sharp} R.,  {Barstow} M.~A.,  {Casewell} S.~L.,    {Hubeny} I.,  2009,
  MNRAS, 395, 2248

\bibitem[\protect\citeauthoryear{{Dobbie}, {Napiwotzki}, {Lodieu}, {Burleigh},
  {Barstow} \& {Jameson}}{{Dobbie} et~al.}{2006}]{dobbie06b}
{Dobbie} P.~D.,  {Napiwotzki} R.,  {Lodieu} N.,  {Burleigh} M.~R.,  {Barstow}
  M.~A.,    {Jameson} R.~F.,  2006, MNRAS, 373, L45

\bibitem[\protect\citeauthoryear{{Dobbie}, {Pinfield}, {Napiwotzki}, {Hambly},
  {Burleigh}, {Barstow}, {Jameson} \& {Hubeny}}{{Dobbie}
  et~al.}{2004}]{dobbie04}
{Dobbie} P.~D.,  {Pinfield} D.~J.,  {Napiwotzki} R.,  {Hambly} N.~C.,
  {Burleigh} M.~R.,  {Barstow} M.~A.,  {Jameson} R.~F.,    {Hubeny} I.,  2004,
  MNRAS, 355, L39

\bibitem[\protect\citeauthoryear{{Eldridge} \& {Tout}}{{Eldridge} \&
  {Tout}}{2004}]{eldridge04}
{Eldridge} J.~J.,  {Tout} C.~A.,  2004, Mem. Soc. Astron. Ital., 75, 694

\bibitem[\protect\citeauthoryear{{Emerson}, {McPherson} \&
  {Sutherland}}{{Emerson} et~al.}{2006}]{emerson06}
{Emerson} J.,  {McPherson} A.,    {Sutherland} W.,  2006, The Messenger, 126,
  41

\bibitem[\protect\citeauthoryear{{Fernandez} \& {Salgado}}{{Fernandez} \&
  {Salgado}}{1980}]{fernandez80}
{Fernandez} J.~A.,  {Salgado} C.~W.,  1980, A\&AS, 39, 11

\bibitem[\protect\citeauthoryear{{Ferrario}, {Wickramasinghe}, {Liebert} \&
  {Williams}}{{Ferrario} et~al.}{2005}]{ferrario05}
{Ferrario} L.,  {Wickramasinghe} D.,  {Liebert} J.,    {Williams} K.~A.,  2005,
  MNRAS, 361, 1131

\bibitem[\protect\citeauthoryear{{Fitzpatrick}}{{Fitzpatrick}}{1999}]{fitzpatr%
ick99}
{Fitzpatrick} E.~L.,  1999, PASP, 111, 63

\bibitem[\protect\citeauthoryear{{Fontaine}, {Brassard} \&
  {Bergeron}}{{Fontaine} et~al.}{2001}]{fontaine01}
{Fontaine} G.,  {Brassard} P.,    {Bergeron} P.,  2001, PASP, 113, 409

\bibitem[\protect\citeauthoryear{{Garcia-Berro}, {Ritossa} \& {Iben}
  Jr.}{{Garcia-Berro} et~al.}{1997}]{garcia97}
{Garcia-Berro} E.,  {Ritossa} C.,    {Iben} Jr. I.,  1997, ApJ, 485, 765

\bibitem[\protect\citeauthoryear{{Gil-Pons}, {Garc{\'{\i}}a-Berro}, {Jos{\'e}},
  {Hernanz} \& {Truran}}{{Gil-Pons} et~al.}{2003}]{gilpons03}
{Gil-Pons} P.,  {Garc{\'{\i}}a-Berro} E.,  {Jos{\'e}} J.,  {Hernanz} M.,
  {Truran} J.~W.,  2003, A\&A, 407, 1021

\bibitem[\protect\citeauthoryear{{Gil-Pons}, {Guti{\'e}rrez} \&
  {Garc{\'{\i}}a-Berro}}{{Gil-Pons} et~al.}{2007}]{gilpons07}
{Gil-Pons} P.,  {Guti{\'e}rrez} J.,    {Garc{\'{\i}}a-Berro} E.,  2007, A\&A,
  464, 667

\bibitem[\protect\citeauthoryear{{Girardi}, {Bressan}, {Bertelli} \&
  {Chiosi}}{{Girardi} et~al.}{2000}]{girardi00}
{Girardi} L.,  {Bressan} A.,  {Bertelli} G.,    {Chiosi} C.,  2000, A\&AS, 141,
  371

\bibitem[\protect\citeauthoryear{{Hambly}, {Hawkins} \& {Jameson}}{{Hambly}
  et~al.}{1993}]{hambly93}
{Hambly} N.~C.,  {Hawkins} M.~R.~S.,    {Jameson} R.~F.,  1993, A\&AS, 100, 607

\bibitem[\protect\citeauthoryear{{Harris}, {Fitzgerald}, {Mehta} \&
  {Reed}}{{Harris} et~al.}{1993}]{harris93}
{Harris} G.~L.~H.,  {Fitzgerald} M.~P.~V.,  {Mehta} S.,    {Reed} B.~C.,  1993,
  AJ, 106, 1533

\bibitem[\protect\citeauthoryear{{Holberg} \& {Bergeron}}{{Holberg} \&
  {Bergeron}}{2006}]{holberg06}
{Holberg} J.~B.,  {Bergeron} P.,  2006, AJ, 132, 1221

\bibitem[\protect\citeauthoryear{{Iben} Jr. \& {Tutukov}}{{Iben} \&
  {Tutukov}}{1985}]{iben85}
{Iben} Jr. I.,  {Tutukov} A.~V.,  1985, ApJS, 58, 661

\bibitem[\protect\citeauthoryear{{Irwin} \& {Lewis}}{{Irwin} \&
  {Lewis}}{2001}]{irwin01}
{Irwin} M.,  {Lewis} J.,  2001, New Astronomy Review, 45, 105

\bibitem[\protect\citeauthoryear{{Jeffries}, {Thurston} \& {Hambly}}{{Jeffries}
  et~al.}{2001}]{jeffries01b}
{Jeffries} R.~D.,  {Thurston} M.~R.,    {Hambly} N.~C.,  2001, A\&A, 375, 863

\bibitem[\protect\citeauthoryear{{Kalirai}, {Fahlman}, {Richer} \&
  {Ventura}}{{Kalirai} et~al.}{2003}]{kalirai03}
{Kalirai} J.~S.,  {Fahlman} G.~G.,  {Richer} H.~B.,    {Ventura} P.,  2003, AJ,
  126, 1402

\bibitem[\protect\citeauthoryear{{Kalirai}, {Hansen}, {Kelson}, {Reitzel},
  {Rich} \& {Richer}}{{Kalirai} et~al.}{2008}]{kalirai08}
{Kalirai} J.~S.,  {Hansen} B.~M.~S.,  {Kelson} D.~D.,  {Reitzel} D.~B.,  {Rich}
  R.~M.,    {Richer} H.~B.,  2008, ApJ, 676, 594

\bibitem[\protect\citeauthoryear{{Kalirai}, {Richer}, {Reitzel}, {Hansen},
  {Rich}, {Fahlman}, {Gibson} \& {von Hippel}}{{Kalirai}
  et~al.}{2005}]{kalirai05}
{Kalirai} J.~S.,  {Richer} H.~B.,  {Reitzel} D.,  {Hansen} B.~M.~S.,  {Rich}
  R.~M.,  {Fahlman} G.~G.,  {Gibson} B.~K.,    {von Hippel} T.,  2005, ApJL,
  618, L123

\bibitem[\protect\citeauthoryear{{Kharchenko}, {Piskunov}, {R{\"o}ser},
  {Schilbach} \& {Scholz}}{{Kharchenko} et~al.}{2005}]{kharchenko05a}
{Kharchenko} N.~V.,  {Piskunov} A.~E.,  {R{\"o}ser} S.,  {Schilbach} E.,
  {Scholz} R.-D.,  2005, A\&A, 438, 1163

\bibitem[\protect\citeauthoryear{{Koester} \& {Reimers}}{{Koester} \&
  {Reimers}}{1981}]{koester81}
{Koester} D.,  {Reimers} D.,  1981, A\&A, 99, L8

\bibitem[\protect\citeauthoryear{{Koester} \& {Reimers}}{{Koester} \&
  {Reimers}}{1993}]{koester93}
{Koester} D.,  {Reimers} D.,  1993, A\&A, 275, 479

\bibitem[\protect\citeauthoryear{{Koester} \& {Reimers}}{{Koester} \&
  {Reimers}}{1996}]{koester96}
{Koester} D.,  {Reimers} D.,  1996, A\&A, 313, 810

\bibitem[\protect\citeauthoryear{{Lampton}, {Margon} \& {Bowyer}}{{Lampton}
  et~al.}{1976}]{lampton76}
{Lampton} M.,  {Margon} B.,    {Bowyer} S.,  1976, ApJ, 208, 177

\bibitem[\protect\citeauthoryear{{Liebert}, {Bergeron} \& {Holberg}}{{Liebert}
  et~al.}{2005}]{liebert05a}
{Liebert} J.,  {Bergeron} P.,    {Holberg} J.~B.,  2005, ApJS, 156, 47

\bibitem[\protect\citeauthoryear{{Liebert}, {Young}, {Arnett}, {Holberg} \&
  {Williams}}{{Liebert} et~al.}{2005}]{liebert05b}
{Liebert} J.,  {Young} P.~A.,  {Arnett} D.,  {Holberg} J.~B.,    {Williams}
  K.~A.,  2005, ApJL, 630, L69

\bibitem[\protect\citeauthoryear{{Marigo} \& {Girardi}}{{Marigo} \&
  {Girardi}}{2007}]{marigo07}
{Marigo} P.,  {Girardi} L.,  2007, A\&A, 469, 239

\bibitem[\protect\citeauthoryear{{Meynet}, {Mermilliod} \& {Maeder}}{{Meynet}
  et~al.}{1993}]{meynet93}
{Meynet} G.,  {Mermilliod} J.,    {Maeder} A.,  1993, A\&AS, 98, 477

\bibitem[\protect\citeauthoryear{{Napiwotzki}, {Green} \&
  {Saffer}}{{Napiwotzki} et~al.}{1999}]{napiwotzki99}
{Napiwotzki} R.,  {Green} P.~J.,    {Saffer} R.~A.,  1999, ApJ, 517, 399

\bibitem[\protect\citeauthoryear{{Nomoto}}{{Nomoto}}{1984}]{nomoto84}
{Nomoto} K.,  1984, ApJ, 277, 791

\bibitem[\protect\citeauthoryear{{Nomoto}}{{Nomoto}}{1987}]{nomoto87}
{Nomoto} K.,  1987, ApJ, 322, 206

\bibitem[\protect\citeauthoryear{{Pietrinferni}, {Cassisi}, {Salaris} \&
  {Castelli}}{{Pietrinferni} et~al.}{2004}]{pietrinferni04}
{Pietrinferni} A.,  {Cassisi} S.,  {Salaris} M.,    {Castelli} F.,  2004, ApJ,
  612, 168

\bibitem[\protect\citeauthoryear{{Piskunov}, {Schilbach}, {Kharchenko},
  {R{\"o}ser} \& {Scholz}}{{Piskunov} et~al.}{2008}]{piskunov08}
{Piskunov} A.~E.,  {Schilbach} E.,  {Kharchenko} N.~V.,  {R{\"o}ser} S.,
  {Scholz} R.-D.,  2008, A\&A, 477, 165

\bibitem[\protect\citeauthoryear{{Poelarends}, {Herwig}, {Langer} \&
  {Heger}}{{Poelarends} et~al.}{2008}]{poelarends08}
{Poelarends} A.~J.~T.,  {Herwig} F.,  {Langer} N.,    {Heger} A.,  2008, ApJ,
  675, 614

\bibitem[\protect\citeauthoryear{{Pye}, {McGale}, {Allan}, {Barber}, {Bertram},
  {Denby}, {Page}, {Ricketts}, {Stewart} \& {West}}{{Pye} et~al.}{1995}]{pye95}
{Pye} J.~P.,  {McGale} P.~A.,  {Allan} D.~J.,  {Barber} C.~R.,  {Bertram} D.,
  {Denby} M.,  {Page} C.~G.,  {Ricketts} M.~J.,  {Stewart} B.~C.,    {West}
  R.~G.,  1995, MNRAS, 274, 1165

\bibitem[\protect\citeauthoryear{{Qian}}{{Qian}}{2008}]{qian08}
{Qian} Y.,  2008, Proc. Astron. Soc. Aust., 25, 36

\bibitem[\protect\citeauthoryear{{Reimers} \& {Koester}}{{Reimers} \&
  {Koester}}{1982}]{reimers82}
{Reimers} D.,  {Koester} D.,  1982, A\&A, 116, 341

\bibitem[\protect\citeauthoryear{{Reimers} \& {Koester}}{{Reimers} \&
  {Koester}}{1989}]{reimers89}
{Reimers} D.,  {Koester} D.,  1989, A\&A, 218, 118

\bibitem[\protect\citeauthoryear{{Ritossa}, {Garcia-Berro} \& {Iben}
  Jr.}{{Ritossa} et~al.}{1996}]{ritossa96}
{Ritossa} C.,  {Garcia-Berro} E.,    {Iben} Jr. I.,  1996, ApJ, 460, 489

\bibitem[\protect\citeauthoryear{{Romanishin} \& {Angel}}{{Romanishin} \&
  {Angel}}{1980}]{romanishin80}
{Romanishin} W.,  {Angel} J.~R.~P.,  1980, ApJ, 235, 992

\bibitem[\protect\citeauthoryear{{Ruiter}, {Belczynski}, {Sim}, {Hillebrandt},
  {Fryer}, {Fink} \& {Kromer}}{{Ruiter} et~al.}{2011}]{ruiter11}
{Ruiter} A.~J.,  {Belczynski} K.,  {Sim} S.~A.,  {Hillebrandt} W.,  {Fryer}
  C.~L.,  {Fink} M.,    {Kromer} M.,  2011, MNRAS, pp 1282--+

\bibitem[\protect\citeauthoryear{{Salaris}, {Serenelli}, {Weiss} \& {Miller
  Bertolami}}{{Salaris} et~al.}{2009}]{salaris09}
{Salaris} M.,  {Serenelli} A.,  {Weiss} A.,    {Miller Bertolami} M.,  2009,
  ApJ, 692, 1013

\bibitem[\protect\citeauthoryear{{Salpeter}}{{Salpeter}}{1955}]{salpeter55}
{Salpeter} E.~E.,  1955, ApJ, 121, 161

\bibitem[\protect\citeauthoryear{{Scalo} \& {Elmegreen}}{{Scalo} \&
  {Elmegreen}}{2004}]{scalo04}
{Scalo} J.,  {Elmegreen} B.~G.,  2004, ARA\&A, 42, 275

\bibitem[\protect\citeauthoryear{{Shafer}}{{Shafer}}{1991}]{shafer91}
{Shafer} R.~A.,  1991, ESA TM-09

\bibitem[\protect\citeauthoryear{{Sharma}, {Pandey}, {Ogura}, {Mito},
  {Tarusawa} \& {Sagar}}{{Sharma} et~al.}{2006}]{sharma06}
{Sharma} S.,  {Pandey} A.~K.,  {Ogura} K.,  {Mito} H.,  {Tarusawa} K.,
  {Sagar} R.,  2006, AJ, 132, 1669

\bibitem[\protect\citeauthoryear{{Siess}}{{Siess}}{2006}]{siess06}
{Siess} L.,  2006, A\&A, 448, 717

\bibitem[\protect\citeauthoryear{{Siess}}{{Siess}}{2007}]{siess07}
{Siess} L.,  2007, A\&A, 476, 893

\bibitem[\protect\citeauthoryear{{Siess}}{{Siess}}{2010}]{siess10}
{Siess} L.,  2010, A\&A, 512, A10+

\bibitem[\protect\citeauthoryear{{Sung}, {Bessell}, {Lee} \& {Lee}}{{Sung}
  et~al.}{2002}]{sung02}
{Sung} H.,  {Bessell} M.~S.,  {Lee} B.,    {Lee} S.,  2002, AJ, 123, 290

\bibitem[\protect\citeauthoryear{{Terndrup}, {Pinsonneault}, {Jeffries},
  {Ford}, {Stauffer} \& {Sills}}{{Terndrup} et~al.}{2002}]{terndrup02}
{Terndrup} D.~M.,  {Pinsonneault} M.,  {Jeffries} R.~D.,  {Ford} A.,
  {Stauffer} J.~R.,    {Sills} A.,  2002, ApJ, 576, 950

\bibitem[\protect\citeauthoryear{{van Dokkum}}{{van
  Dokkum}}{2001}]{vandokkum01}
{van Dokkum} P.~G.,  2001, PASP, 113, 1420

\bibitem[\protect\citeauthoryear{{van Leeuwen}}{{van
  Leeuwen}}{2009}]{vanleeuwen09}
{van Leeuwen} F.,  2009, A\&A, 497, 209

\bibitem[\protect\citeauthoryear{{Vennes} \& {Kawka}}{{Vennes} \&
  {Kawka}}{2008}]{vennes08}
{Vennes} S.,  {Kawka} A.,  2008, MNRAS, 389, 1367

\bibitem[\protect\citeauthoryear{{Vennes}, {Thejll}, {Galvan} \&
  {Dupuis}}{{Vennes} et~al.}{1997}]{vennes97}
{Vennes} S.,  {Thejll} P.~A.,  {Galvan} R.~G.,    {Dupuis} J.,  1997, ApJ, 480,
  714

\bibitem[\protect\citeauthoryear{{Wachter}, {Schr{\"o}der}, {Winters}, {Arndt}
  \& {Sedlmayr}}{{Wachter} et~al.}{2002}]{wachter02}
{Wachter} A.,  {Schr{\"o}der} K.-P.,  {Winters} J.~M.,  {Arndt} T.~U.,
  {Sedlmayr} E.,  2002, A\&A, 384, 452

\bibitem[\protect\citeauthoryear{{Wagenhuber} \& {Groenewegen}}{{Wagenhuber} \&
  {Groenewegen}}{1998}]{wagenhuber98}
{Wagenhuber} J.,  {Groenewegen} M.~A.~T.,  1998, A\&A, 340, 183

\bibitem[\protect\citeauthoryear{{Wagoner}, {Fowler} \& {Hoyle}}{{Wagoner}
  et~al.}{1967}]{wagoner67}
{Wagoner} R.~V.,  {Fowler} W.~A.,    {Hoyle} F.,  1967, ApJ, 148, 3

\bibitem[\protect\citeauthoryear{{Wanajo}, {Nomoto}, {Janka}, {Kitaura} \&
  {M{\"u}ller}}{{Wanajo} et~al.}{2009}]{wanajo09}
{Wanajo} S.,  {Nomoto} K.,  {Janka} H.,  {Kitaura} F.~S.,    {M{\"u}ller} B.,
  2009, ApJ, 695, 208

\bibitem[\protect\citeauthoryear{{Weidemann}}{{Weidemann}}{2005}]{weidemann05}
{Weidemann} V.,  2005, in {D.~Koester \& S.~Moehler} ed., 14th European
  Workshop on White Dwarfs Vol.~334 of Astronomical Society of the Pacific
  Conference Series, {On Supermassive White Dwarfs}.
pp 15--+

\bibitem[\protect\citeauthoryear{{Williams}}{{Williams}}{2004}]{williams04a}
{Williams} K.~A.,  2004, ApJ, 601, 1067

\bibitem[\protect\citeauthoryear{{Williams}, {Bolte} \& {Koester}}{{Williams}
  et~al.}{2004}]{williams04b}
{Williams} K.~A.,  {Bolte} M.,    {Koester} D.,  2004, ApJL, 615, L49

\bibitem[\protect\citeauthoryear{{Williams}, {Bolte} \& {Koester}}{{Williams}
  et~al.}{2009}]{williams09}
{Williams} K.~A.,  {Bolte} M.,    {Koester} D.,  2009, ApJ, 693, 355

\bibitem[\protect\citeauthoryear{{Woo}, {Gallart}, {Demarque}, {Yi} \&
  {Zoccali}}{{Woo} et~al.}{2003}]{woo03}
{Woo} J.-H.,  {Gallart} C.,  {Demarque} P.,  {Yi} S.,    {Zoccali} M.,  2003,
  AJ, 125, 754

\bibitem[\protect\citeauthoryear{{Yuan}}{{Yuan}}{1992}]{yuan92}
{Yuan} J.~W.,  1992, A\&A, 261, 105

\bibitem[\protect\citeauthoryear{{Yungelson} \& {Livio}}{{Yungelson} \&
  {Livio}}{2000}]{yungelson00}
{Yungelson} L.~R.,  {Livio} M.,  2000, ApJ, 528, 108

\end{thebibliography}

\bsp

\label{lastpage}

\end{document}